\documentclass[a4paper,11pt]{article}
\pdfoutput=1 

\usepackage{jcappub} 

\usepackage[T1]{fontenc} 

\usepackage{graphicx}  
\usepackage{dcolumn}   
\usepackage{bm}        
\usepackage{amssymb}   
\usepackage {url}
\usepackage{amsmath}
\usepackage{textcomp}
\usepackage{hyperref}
\usepackage{multirow}
\usepackage{upgreek}
\usepackage{booktabs}
\usepackage{xfrac}
\usepackage{color}
\usepackage{wrapfig}
\usepackage[normalem]{ulem}
\usepackage{siunitx}
\usepackage{float}

\title{Seasonal Variation of the Underground Cosmic Muon Flux \\Observed at Daya Bay}

\collaboration{The Daya Bay Collaboration}

\def\ECUST{a}
\def\Wisconsin{b}
\def\Yale{c}
\def\BNL{d}
\def\NTU{e}
\def\NUU{f}
\def\NJU{g}
\def\IHEP{h}
\def\CUHK{i}
\def\SDU{j}
\def\TsingHua{k}
\def\NCEPU{l}
\def\SZU{m}
\def\ZSU{n}
\def\Dubna{o}
\def\Siena{p}
\def\UIUC{q}
\def\LBNL{r}
\def\SJTU{s}
\def\BNU{t}
\def\XJTU{u}
\def\BCC{2}
\def\BCCnote{\note{Now at: Department of Chemistry and Chemical Technology, Bronx Community College, Bronx, New York  10453}}
\def\UH{w}
\def\CIAE{x}
\def\VirginiaTech{y}
\def\USTC{z}
\def\NCTU{aa}
\def\NanKai{ab}
\def\UC{ac}
\def\TempleUniversity{ad}
\def\DGUT{ae}
\def\UCB{af}
\def\HKU{ag}
\def\Charles{ah}
\def\IIT{ai}
\def\Princeton{aj}
\def\CUC{ak}
\def\CalTech{al}
\def\WM{am}
\def\RPI{an}
\def\CGNPG{ao}
\def\NUDT{ap}
\def\IowaState{aq}
\def\CQU{ar}
\author[\ECUST]{F.~P.~An}
\author[\Wisconsin]{A.~B.~Balantekin}
\author[\Yale]{H.~R.~Band}
\author[\BNL]{M.~Bishai}
\author[\NTU,\NUU]{S.~Blyth}
\author[\NJU]{D.~Cao}
\author[\IHEP]{G.~F.~Cao}
\author[\IHEP]{J.~Cao}
\author[\CUHK]{Y.~L.~Chan}
\author[\IHEP]{J.~F.~Chang}
% ...

\newcommand{\teff}{T_{\mathrm{eff}}}
\newcommand{\ethr}{E_{\mathrm{thr}}}
\newcommand{\mev}{\,\mathrm{MeV}}

\abstract{
The Daya Bay Experiment consists of eight identically designed detectors located in three underground experimental halls
named as EH1, EH2, EH3, with 250, 265 and 860 meters of water equivalent vertical overburden, respectively.
Cosmic muon events have been recorded over a two-year period. The underground muon rate is observed to be positively correlated with the effective atmospheric temperature and to follow a seasonal modulation
pattern. The correlation coefficient $\alpha$, describing how a variation in the muon rate relates to a variation in the effective atmospheric temperature, is found to be $\alpha_{\text{EH1}} = 0.362\pm0.031$, $\alpha_{\text{EH2}} = 0.433\pm0.038$ 
and $\alpha_{\text{EH3}} = 0.641\pm0.057$ for each experimental hall.}

\begin{document}
\maketitle
\flushbottom

%
%
\section{Introduction}

Early investigations into the nature and origin of cosmic rays included searches for correlations between the penetrating component of the cosmic ray flux and atmospheric variables~\cite{Barrett}.
We now know that this penetrating component is composed of positive and negative muons.  The muons result from the decay of charged mesons produced by interactions of primary cosmic rays with the upper atmosphere.  Furthermore, a number of experiments have observed the underground muon intensity to be positively correlated with the atmospheric temperature~\cite{Barrett, Barrett2, Sherman, Castagnoli, Fenton, Matsushiro, Poatina1, Poatina2, Baksan, Amanda, Macro, Borexino, LVD, Icecube2011, MinosND, MinosFD, Gerda, Utah, DC}. As the temperature increases, the atmosphere becomes less dense,  and the probability for a meson to interact with molecules in the atmosphere is reduced. The corresponding increase in meson decays yields a larger muon intensity over the summer months.  The great majority of the experimental results are reported in terms of $\alpha$, the correlation coefficient between the muon flux and the atmospheric temperature. This coefficient increases as a function of overburden, and hence Daya Bay, with three underground experimental halls at different depths, is an ideal setup to perform such a measurement.

%
%
\section{Daya Bay Experiment}
\begin{wrapfigure}{l}{0.5\textwidth}
\begin{center}
\includegraphics[width=\linewidth]{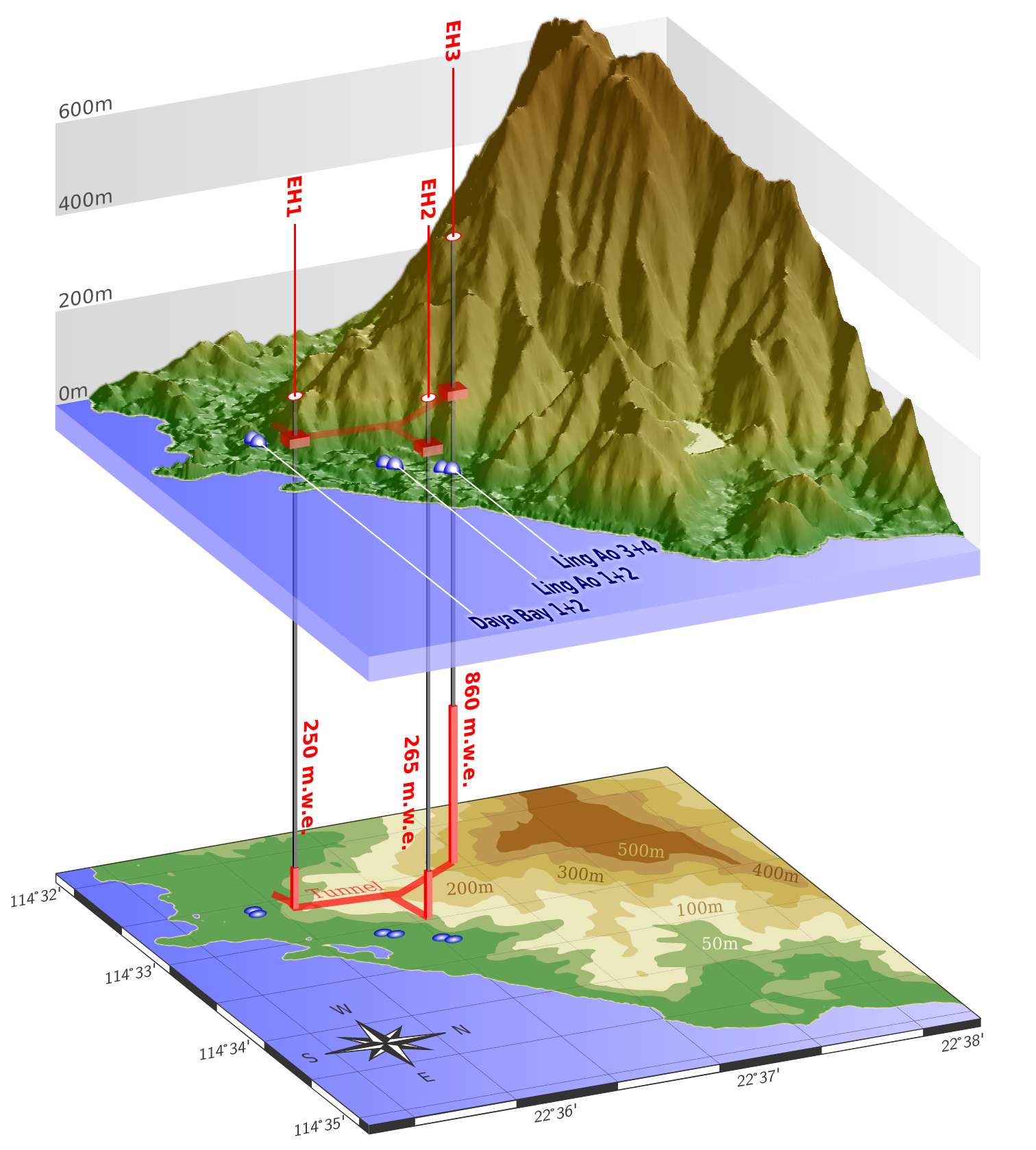}
\end{center}
\caption{\label{fig:overburden}
Location of the Nuclear Power Plants and of the Experimental halls,
together with the elevation profile of the mountain above the
experimental halls.}
\end{wrapfigure}

Daya Bay is designed to measure the previously unknown value of the neutrino mixing angle $\theta_{13}$ by measuring the survival probability of electron antineutrinos from nuclear reactors at suitable distances. Electron antineutrinos are detected via inverse beta decay
(IBD) $\bar{\nu_e}p\to e^+ n$ where the positron energy is less than 10 MeV for reactor antineutrinos.
Figure~\ref{fig:overburden} shows a diagram of the Daya Bay experimental site.  The Daya Bay Nuclear Power Plant complex consists of six reactors.  The experiment uses eight identically designed antineutrino detectors (ADs) located in three underground experimental halls (EHs). Two halls, EH1 and EH2, are located near the reactor cores, while the last hall, EH3,
is located farther away at a distance optimized to measure $\theta_{13}$ through neutrino oscillation.  Daya Bay
began operations in December 2011 with six ADs: two in EH1, one in EH2, and three in EH3.
In Summer 2012, the remaining two ADs were installed, one in EH2 and the other one in EH3, and operations began with all eight ADs.
The vertical overburden and the measured muon flux at each EH are listed in Table~\ref{tab:underground}.

The ADs at each hall are contained inside a muon detector system,
which consists of a two-zone pure water Cherenkov
detector, referred to as the inner and outer water shields
(IWS and OWS), covered on top by an array of resistive
plate chambers (RPCs). The water pool is designed so that at least 2.5~m of water surrounds each AD in every direction.
There are a total of 288 water pool PMTs at each of the near sites (EH1, EH2) and 384 PMTs at EH3, distributed in the inner and outer regions, which are optically separated by Tyvek sheets.
Each AD consists of three
nested cylindrical volumes separated by transparent acrylic vessels.
The inner acrylic vessel (IAV) has a 3.1-m diameter and is filled with
20~tons of 0.1\% gadolinium-doped liquid scintillator (GdLS) as the primary antineutrino target. The 4-m diameter outer acrylic vessel (OAV)
surrounding the target is filled with about 21~tons of undoped liquid scintillator (LS), increasing
the efficiency of detecting gamma rays produced in the GdLS region.
The outermost stainless steel vessel has a diameter of 5~m and
is filled with 37~tons of mineral oil. A total of 192
20-cm photomultiplier tubes (PMTs) are radially positioned
in the mineral-oil region of each AD. Further details on the
AD, including details of calibration and vertex reconstruction,
can be found in Ref.~\cite{SideBySideComparison,LongOs}.  The muon system design and performance are described
in detail in Ref.~\cite{muonPaper}.

\begin{table}
\centering
\begin{tabular}{lcc}
\hline
     Hall  &  Overburden        &  Muon flux  \\
             & m \quad mwe       &  Hz/m$^2$   \\
    \hline
    EH1         & \phantom{0}93 \quad 250 &  $1.16  \pm 0.11$                   \\
    EH2         &           100 \quad 265 &  $0.86  \pm 0.09$                   \\
    EH3         &           324 \quad 860 &  $0.054 \pm 0.006$                   \\
\hline
  \end{tabular}
  \caption{ \label{tab:underground}Vertical overburden and measured muon flux at the three experimental halls~\cite{muonPaper}.}
\end{table}

%
%
\section{Muon Data}
\subsection {Muon Event Selection Criteria}
\begin{figure}[tbh]
\centering
\includegraphics[width=0.5\linewidth]{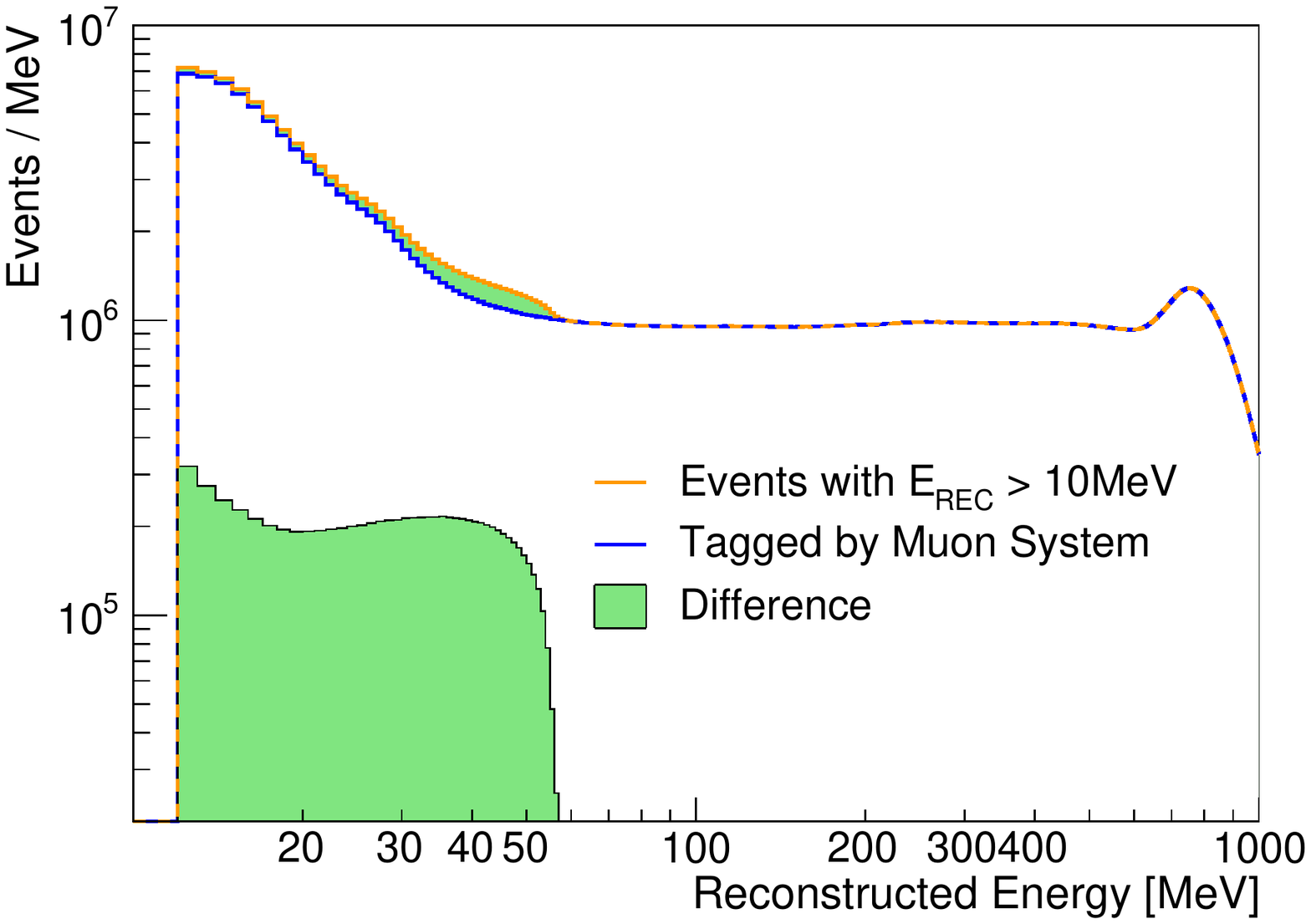}
\caption{\label{fig:michel_electrons}Reconstructed energy spectrum of muon candidate events in EH1 AD1 with and without the muon tag.}
\end{figure}

A muon candidate is defined as an event where (\textsc{i}) the reconstructed energy in an AD is larger than $60\mev$,
and (\textsc{ii}) more than 12 photomultipliers in the muon system (either IWS or OWS) produce a trigger within a
$2 \, \upmu \mathrm{s}$ time window. We refer to (\textsc{ii}) as the muon tag.
The reconstructed energy spectrum of events with energy greater than $10\mev$ is shown in Fig.~\ref{fig:michel_electrons},
where the orange histogram represents all AD events with  energy larger than $10\mev$, and the blue histogram represents the muon system tagged candidates. The difference between the two distributions is highlighted in green, and shows that
untagged events with more than $20\mev$ reconstructed energy
experience a cut-off at roughly half the mass of a muon, typical of electrons originating from stopping muon decay. Such events are
cleanly rejected by our selection criteria.

\begin{figure}[tbh]
\includegraphics[width=0.5\linewidth]{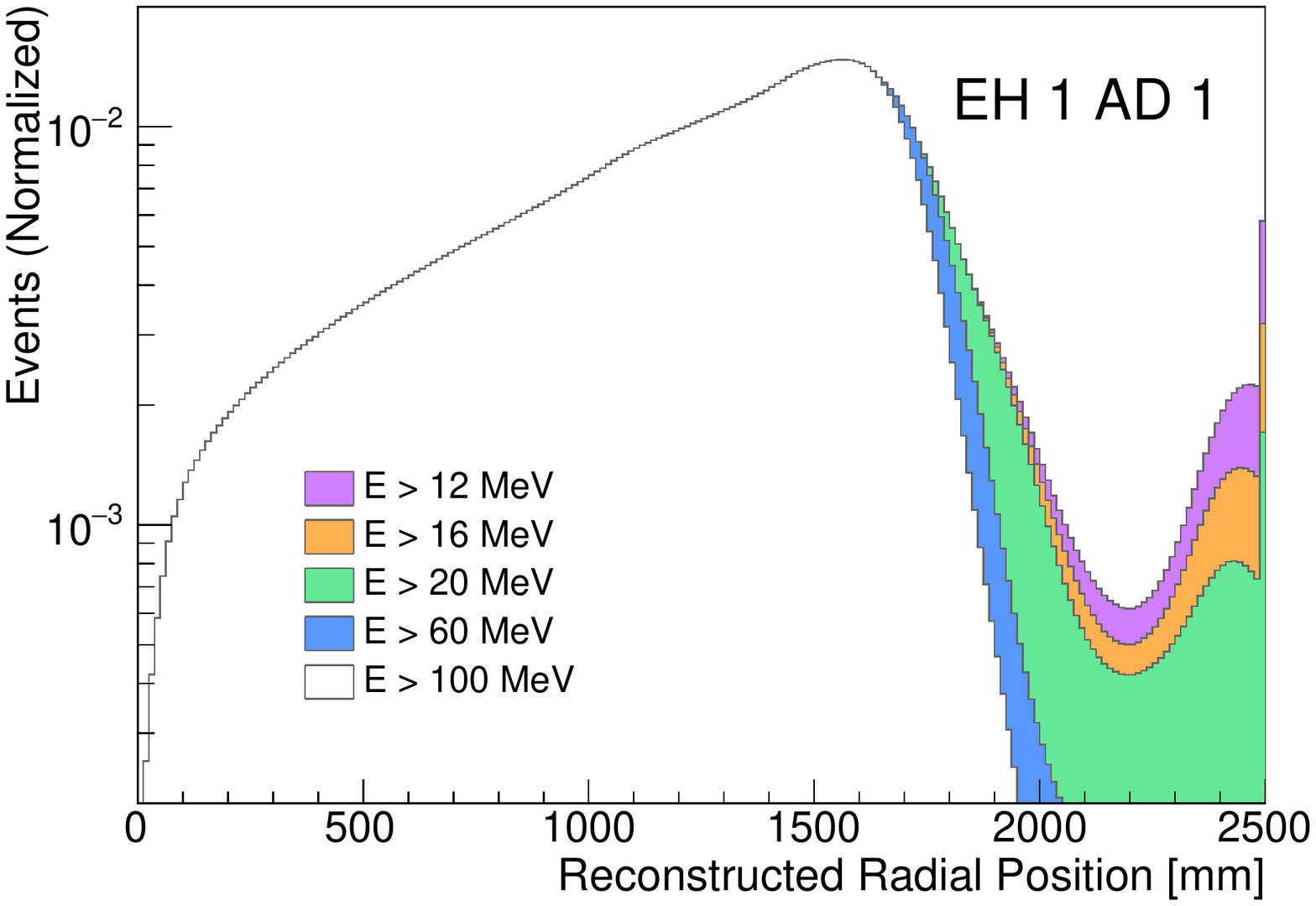}
\hfill
\includegraphics[width=0.5\linewidth]{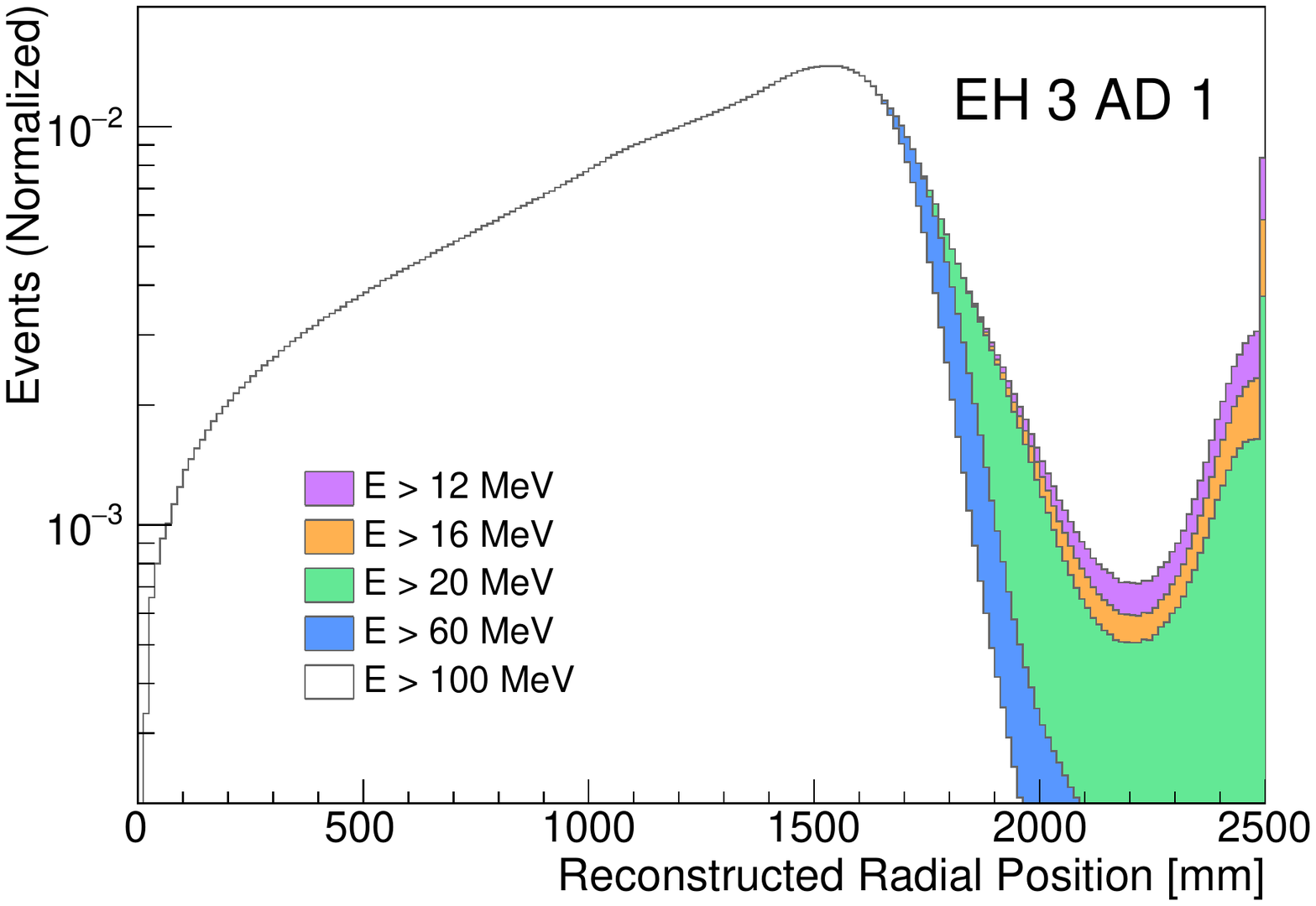}
\caption{\label{fig:mu_candidates}Comparison of the reconstructed radial distribution of muon candidate events in EH1 AD1 (left) and EH3 AD1 (right). In the analysis, the MO region is treated as a single bin [$2000\,$mm, $2500\,$mm]; here we provide finer binning for comparison only.}
\end{figure}

An important requirement for the selection criteria is to be stable over time, which in turn ensures the stability of the muon tagging procedure. The stability of the muon system is described in~\cite{muonPaper}.
The stability of the energy scale for events taking place in both the GdLS and LS regions
is guaranteed by weekly calibration campaigns~\cite{LongOs} that provide the calibration
constants used in the standard IBD analysis. However, the stability of the
events taking place in the external buffer cannot be easily assessed. This region is filled
with mineral oil (MO), and it was designed to shield the target volume from external
radioactivity, hence it lacks any calibration system. Neutrino interactions taking
place in the buffer result in no scintillation light, but muons passing through this region emit
Cherenkov light, potentially resulting 
in events that are uncalibrated and should be vetoed. 
The left panel of Fig.~\ref{fig:mu_candidates} shows
the event vertex distribution as a function of the energy cut for energy depositions taking place in EH1 AD1.
Events whose vertex is reconstructed in the MO buffer ($\mathrm{R}>2\,\mathrm{m}$) clearly cluster
at low energy, and are efficiently rejected by a $60\mev$ cut.
To further enhance the stability over time of the muon selection criteria,
we correct for the permille-level energy drifts that the ADs experience
because of the liquid scintillator aging. 
Such corrections are derived from fits to the $^{208}\mathrm{Tl}$ spectrum due to residual radioactivity in the scintillator. 

The only exception to the selection criteria introduced so far is EH3 AD1. We know that in this
AD a tiny leak of liquid
scintillator from the LS region to the buffer started in Summer 2012.
The LS and MO levels stabilized in 2014 when an estimated 50 L of LS had leaked into the MO~\cite{An:2015qga}.
As a consequence, the MO light yield increased, and the $60\mev$ energy cut was more likely to select
muons crossing the buffer region. To account for such
a difference in light yield, we raise the energy cut of EH3 AD1 to $100\mev$, and we consider this AD separately
from the others when computing systematic uncertainties.

\subsection{Muon Rate Variation Over Time}
Muon events are selected from a dataset collected between December 2011
and November 2013.
The first 7 months of data taking are characterised by having only 6 operating ADs~\cite{An:2013uza},
while the last 13 months exploit the full 8-AD configuration.

\begin{figure}[tbh]
\includegraphics[width=\textwidth]{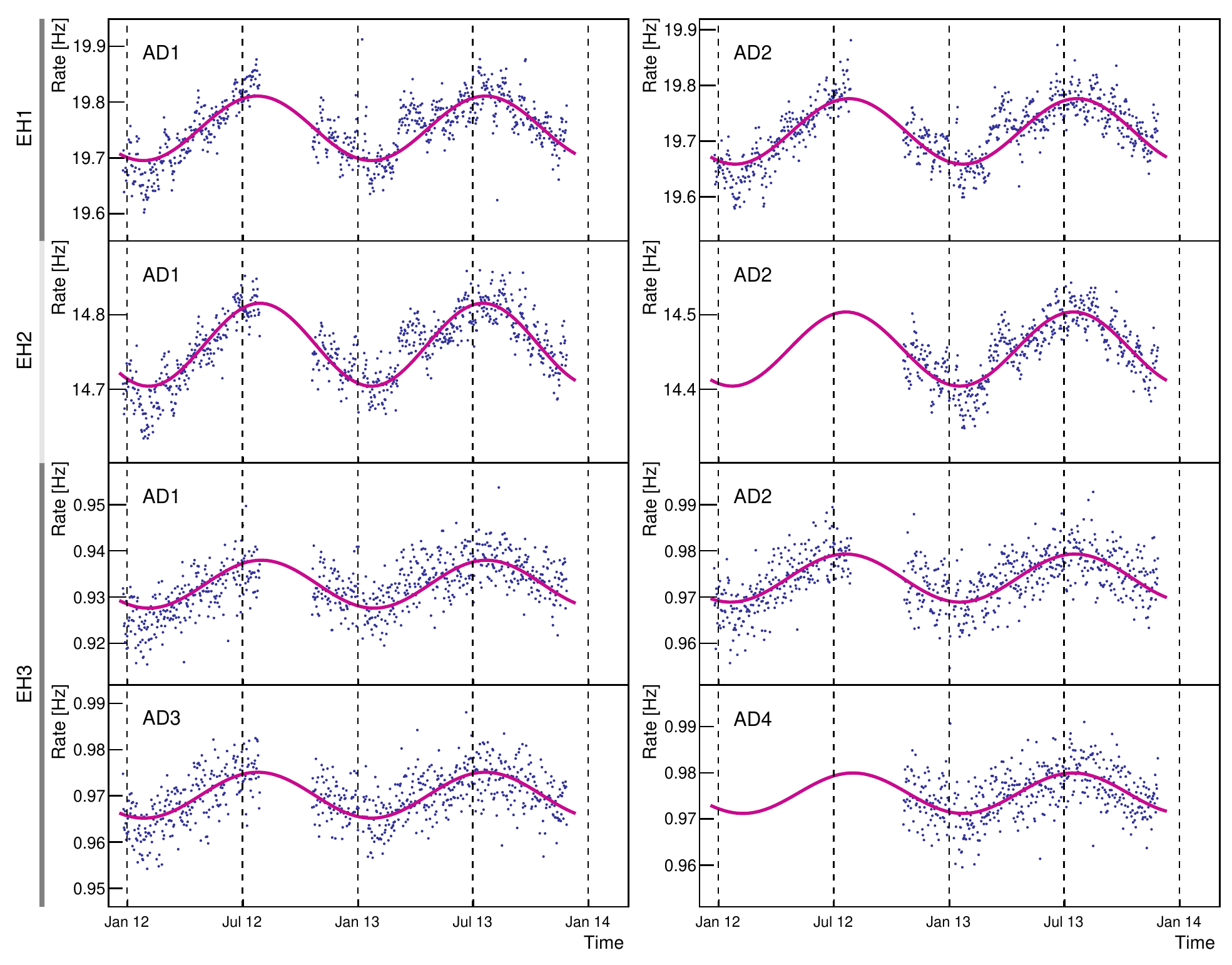}
\caption{Daily-binned muon rate as function of time in the eight ADs. The solid line shows the result of a sinusoidal fit to data. Fit parameters are shown in
Table~\ref{tab:muon_fit_params}.}\label{fig:fitted_muon_rate}
\end{figure}

The daily muon rate as a function of time in all the ADs is shown in Fig.~\ref{fig:fitted_muon_rate}. A fit with a sinusoidal function
is performed to each  AD separately, with the aim of checking if the modulation features are compatible among ADs.
Fit parameters are reported in Table~\ref{tab:muon_fit_params}.
The oscillation period is compatible with one solar year, and the position of the oscillation maximum
(i.e. the oscillation phase) occurs  consistently towards the end of July.
The oscillation amplitude depends on the average muon energy, and therefore on the overburden. The average rate of EH3 AD1 is lower than the other ADs in the same experimental
hall because of the tighter energy cut.

We stress that the sinusoidal fit is not used  in the correlation analysis, but merely to enable a comparison among ADs as described above. 
The inability of a simple sinusoid to describe features such as the jump in the muon rate around March 2013 
should not be  considered a limitation. Indeed,
the daily correlation of muon data with atmospheric temperature data is able to correctly account for such deviations from the sinusoidal function.

\begin{table}
\centering
\begin{tabular}{llll}
\multicolumn{2}{c}{Detector} & Maximum & Period \\
\hline
\rule{0pt}{3ex}
\multirow{2}{*}{EH1} & AD1 & $21 \, \mathrm{Jul} \pm 2  \, \mathrm{days} $ & $361 \pm 1 \, \mathrm{day} $ \\
                     & AD2 & $21 \, \mathrm{Jul} \pm 2  \, \mathrm{days} $ & $361 \pm 1 \, \mathrm{day} $ \\
\rule{0pt}{3ex}
\multirow{2}{*}{EH2} & AD1 & $17 \, \mathrm{Jul} \pm 2  \, \mathrm{days} $ & $353 \pm 1 \, \mathrm{day} $ \\
                     & AD2 & $15 \, \mathrm{Jul} \pm 4  \, \mathrm{days} $ & $360 \pm 3 \, \mathrm{days} $ \\
\rule{0pt}{3ex}
\multirow{4}{*}{EH3} & AD1 & $22 \, \mathrm{Jul} \pm 4  \, \mathrm{days} $ & $356 \pm 3 \, \mathrm{days} $ \\
                     & AD2 & $18 \, \mathrm{Jul} \pm 5  \, \mathrm{days} $ & $363 \pm 4 \, \mathrm{days} $ \\
                     & AD3 & $20 \, \mathrm{Jul} \pm 5  \, \mathrm{days} $ & $360 \pm 4 \, \mathrm{days} $ \\
                     & AD4 & $15 \, \mathrm{Jul} \pm 10 \, \mathrm{days} $ & $348 \pm 8 \, \mathrm{days} $ \\
\end{tabular}
\caption{\label{tab:muon_fit_params}Parameters resulting from fitting the muon modulation
with a sinusoidal function.}
\end{table}

\subsection {Muon Threshold Energy}
\label{sec:e_thr}

The intensity of the cosmic muon flux  $I_{\mu}$ is known to be dependent on the muon energy
$E_{\mu}$~\cite{PDG2014_Undergroud_Muon}.
In the case of underground experiments, the muon energy spectrum is truncated, because low-energy muons
get stopped by the rock above each experimental site. Here we define ``threshold energy'' ($\ethr$) to be the
minimum energy that a muon must have at the surface in order to reach an experimental hall.
Liquid scintillator experiments are in general not able to measure $E_{\mu}$, hence the muon rate they measure
is the integral of $I_{_\mu}(E_{\mu})$ from $\ethr$ to the maximum cosmic muon energy.
As a consequence, $\ethr$ is one of the most important parameters differentiating experiments that perform
inclusive measurements of the muon flux at different underground depths.
In this analysis, $\ethr$ plays two roles: first, it is involved in the procedure to compute the effective atmospheric temperature $\teff$
(Eq.~\ref{eq:weights}), and second, it allows us to compare the measured muon modulation against both an atmospheric model
(Eq.~\ref{eq:theoretical_alpha1}) and other experiments (Fig.~\ref{fig:comparison_other_exp}).

\begin{figure}[tbh]
\centering
\includegraphics[width=0.49\linewidth]{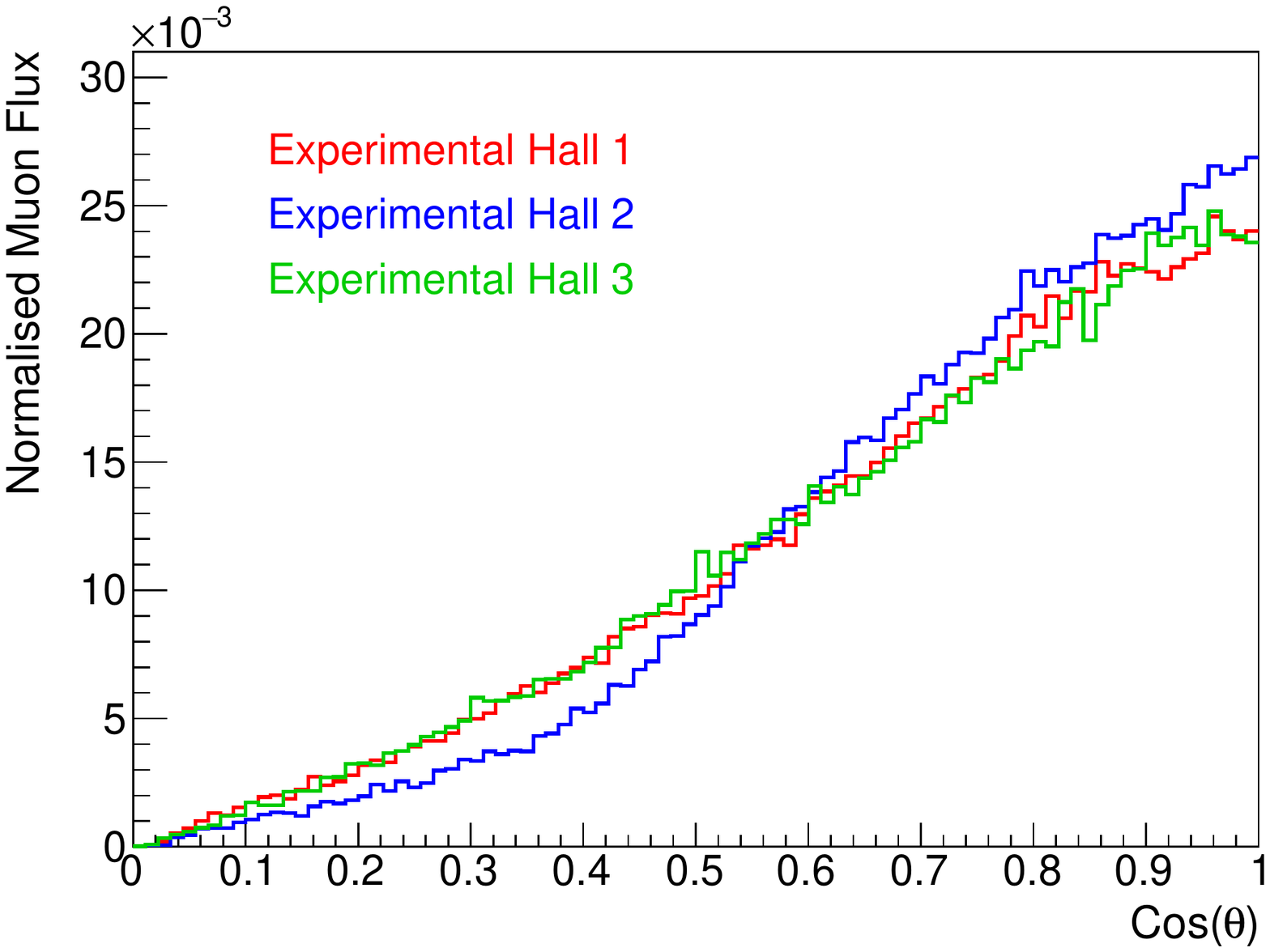} \hfill
\includegraphics[width=0.49\linewidth]{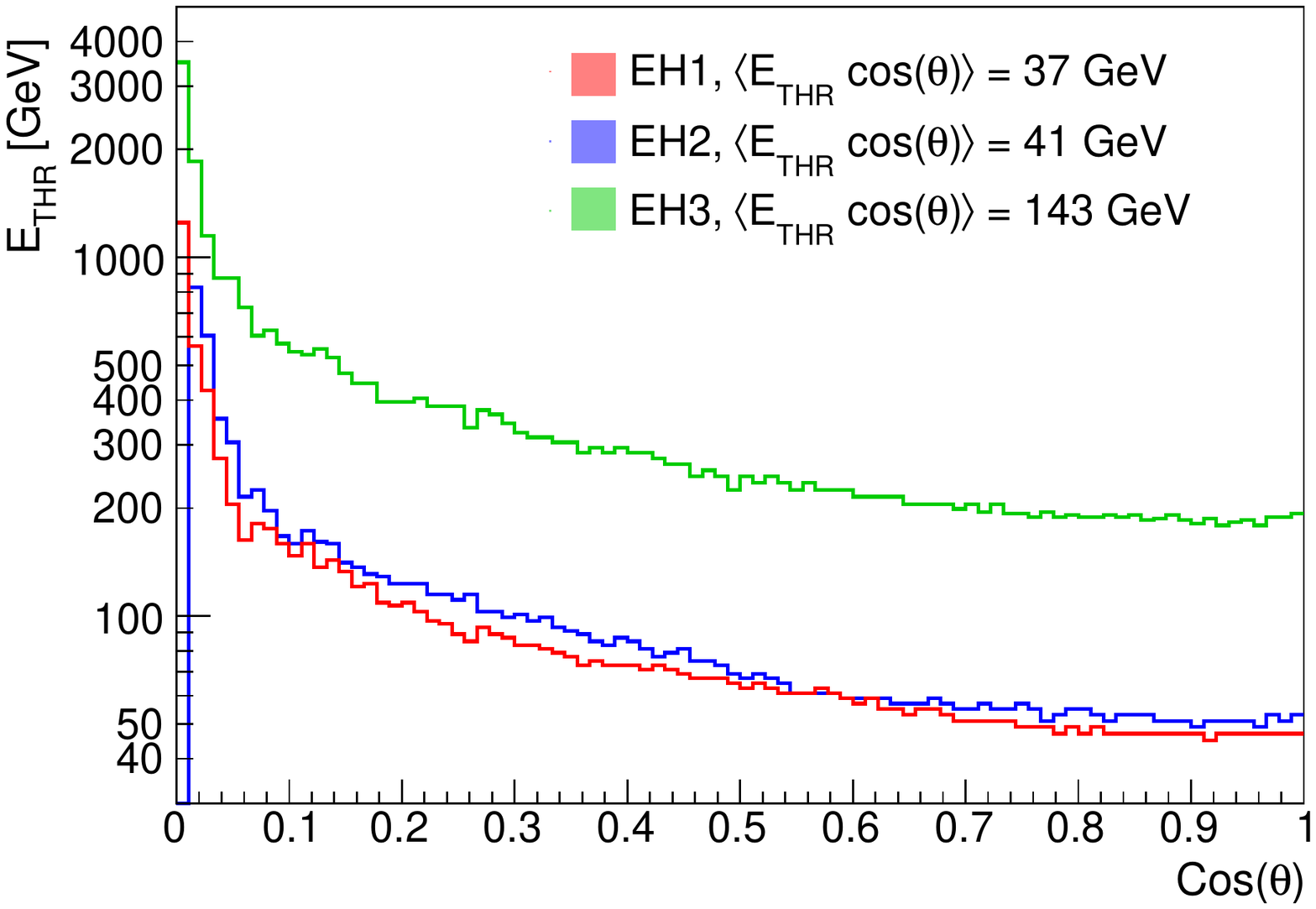}
\caption{\label{fig:threshold_energy}(Left) Underground muon flux normalised to unity as a function of the cosine of the zenith angle $\cos \theta$ at the surface. 
(Right) Minimum or threshold energy needed for a muon to reach an experimental hall
$\ethr$ as a function of  $\cos\theta$. Small-scale structures are ascribable to statistical fluctuations of the MC simulation and not 
to a rapidly changing mountain profile.}
\end{figure}

To determine $\ethr$ at the three Daya Bay experimental sites we rely on MC simulations.
Namely, we simulate cosmic muon propagation through the  overburden
using the topographic maps of the three experimental sites (see Fig.~\ref{fig:overburden}) with the MUSIC simulation package~\cite{Music}.
The overburden is assumed to be made of standard rock, defined to have atomic number 11,
atomic mass 22, and density $2.65\,\mathrm{g/cm^3}$.
We generate more than 1 million muons at the surface of each site according to the modified
Gaisser formula \cite{ImprovedGaisser}, where the exact number is chosen such that $10^5$ muons reach the experimental hall.
For each muon, we record its zenith angle $\theta$, its azimuthal angle $\phi$ and its energy, both at the
surface (before propagation) and at experimental hall (after propagation), which results in a
site-dependent $E_{\mu}(\theta , \phi)$ distribution. 
In the theoretical model describing the seasonal modulation of the muon flux, $\ethr$ always appears
 multiplied by the cosine of the zenith angle (evaluated at the surface), resulting in the expression $\ethr \cos \theta$.
We therefore marginalise the azimuthal dependency,
$E_{\mu}(\theta) = \int \mathrm{d} \phi \, E_{\mu}(\theta , \phi)$, and bin the resulting energy distribution in terms of $\cos \theta$. Each angular bin is then characterised
by an energy spectrum, whose starting point we define to be $\ethr$. This procedure results in 
a $\cos\theta$- and site-dependent  $\ethr$, which is shown in the right panel of Fig.~\ref{fig:threshold_energy}.

The zenith angle dependency of the term $\ethr \cos \theta$ results from two competing effects:
(\textsc{i}) cosmic ray primaries coming from the horizon interact higher in the atmosphere, enhancing the probability
of secondary mesons to decay into muons, and (\textsc{ii}) a lower fraction of such cosmic muons reach the detector
because the overburden increases rapidly towards the horizon.
In principle, the former considerations could be exploited to investigate how the muon rate
modulation changes as a function of the zenith angle~\cite{MinosND}. However,
pointing information is not currently extracted in our reconstruction, so this analysis considers
only inclusive quantities.
For this reason, we average $\ethr \cos \theta$ over all the zenith angles:
\begin{equation}
 \langle \ethr \cos \theta \rangle = \sum_{i} \ethr \left( c_{i} \right) \cdot
n\left(  c_{i} \right) \cdot   c_{i}
\end{equation}
where the index $i$ runs over the bins of the histograms shown in
Fig.~\ref{fig:threshold_energy}, $c_{i}$ is the value of $\cos \theta$ at the bin center,
and $n(c_i)$ is the muon flux normalized to unity evaluated at
the $i$-th angular bin, as shown in the left panel of Fig.~\ref{fig:threshold_energy}. 
The $\langle \ethr \cos \theta \rangle$ values for EH1, EH2 and EH3 are 
$37\,\mathrm{GeV}$, $41\,\mathrm{GeV}$ and $143\,\mathrm{GeV}$, respectively.

The systematic uncertainty on $\langle \ethr \cos \theta \rangle$ breaks down into three major components:
(\textsc{i}) angular resolution,
(\textsc{ii}) energy resolution, and (\textsc{iii}) imperfect
knowledge of the topographic map.
We quantify (\textsc{i}) as the bin width chosen for the  $\cos \theta$ distribution, which results in a 1.1\% relative
uncertainty, and (\textsc{ii}) as the bin width of the $\ethr$ spectrum, which results in a $2\, \mathrm{GeV}/\ethr$
relative uncertainty. Component (\textsc{iii})
is evaluated by shifting the mountain
elevation profile up and down by $6\,\mathrm{m}$, which is the altitude resolution of the topographic map.
This procedure yields a  $6\%$ relative uncertainty. It is worth stressing that
both an increase in the elevation profile and an increase in the rock density
result in a larger $\ethr$, since both contribute to enhance the mountain stopping power.
The density difference between standard rock and the actual Daya Bay rock is around 2\%, 
hence the elevation uncertainty conservatively covers the density uncertainty.  
%
%
The overall relative systematic uncertainty on $\langle \ethr \cos \theta \rangle$ is assessed by summing the squared values (\textsc{i}), (\textsc{ii}) and (\textsc{iii}), which gives an overall 7\% systematic uncertainty.

%
%
\section{Temperature Data}
\subsection{Effective Atmospheric Temperature }
The atmospheric temperature data at the Daya Bay site was obtained from the ERA-Interim database supplied by the European Centre for Medium-Range Weather Forecasts (ECMWF)~\cite{ecmwf}. The database comprises different types of measurements (ground level, sounding balloon, satellite) at many locations over the world, and exploits a global atmospheric model to interpolate to a particular location~\cite{weather_interpolation}. Our analysis relies on the temperature values computed at the Daya Bay site (22.6$^\circ$N, 114.5$^\circ$E), which are provided four times a day (midnight, 6am, noon, 6pm) at 37 discrete pressure levels ranging from 1 hPa to 1000 hPa.
The interpolated temperature dataset has a spatial resolution of $0.25^{\circ} \times 0.25^{\circ}$, hence all the three Daya Bay experimental halls share the same raw  temperature dataset.

\begin{figure}[tbh]
\includegraphics[width=0.5\linewidth]{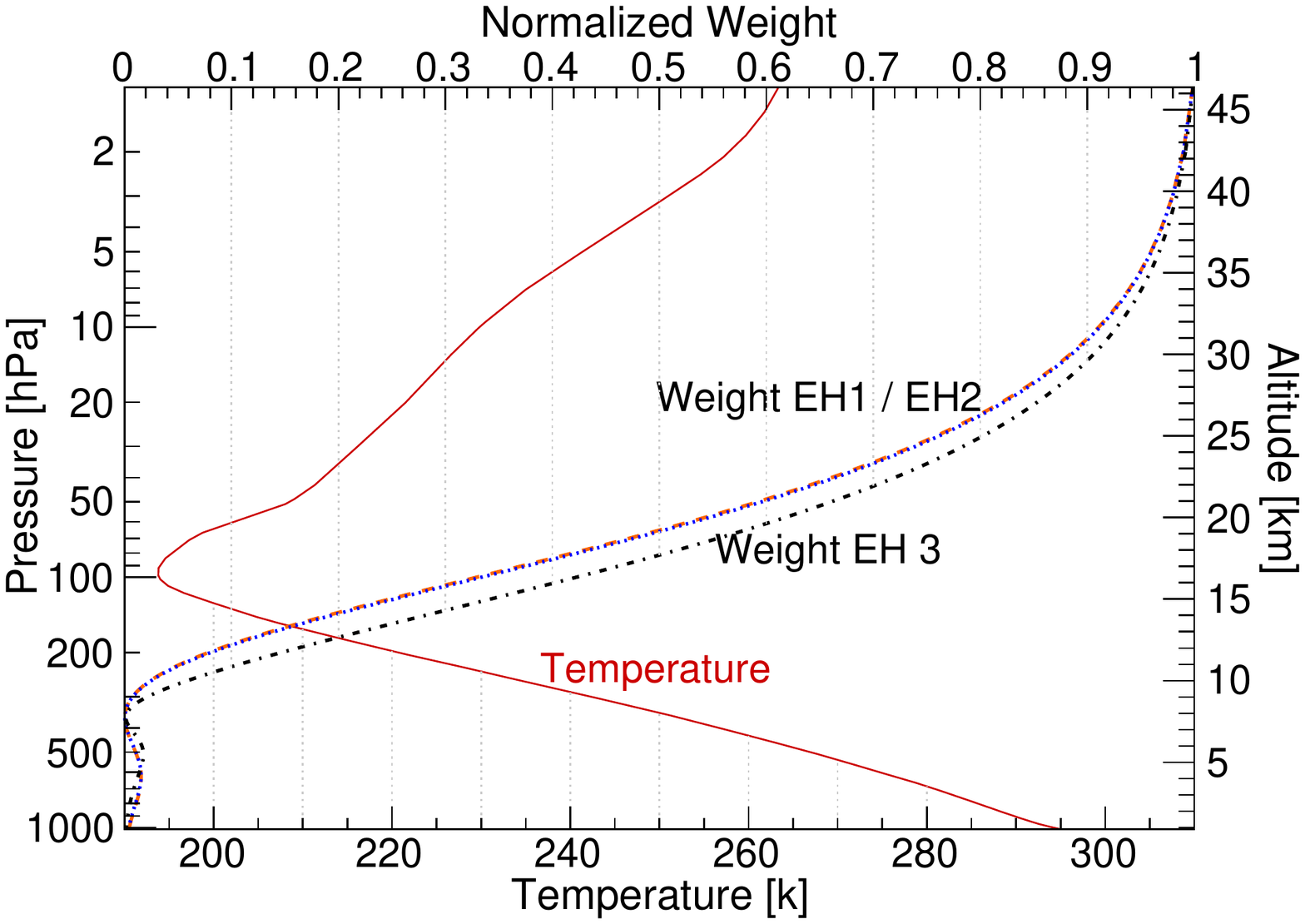} \hfill
\includegraphics[width=0.5\linewidth]{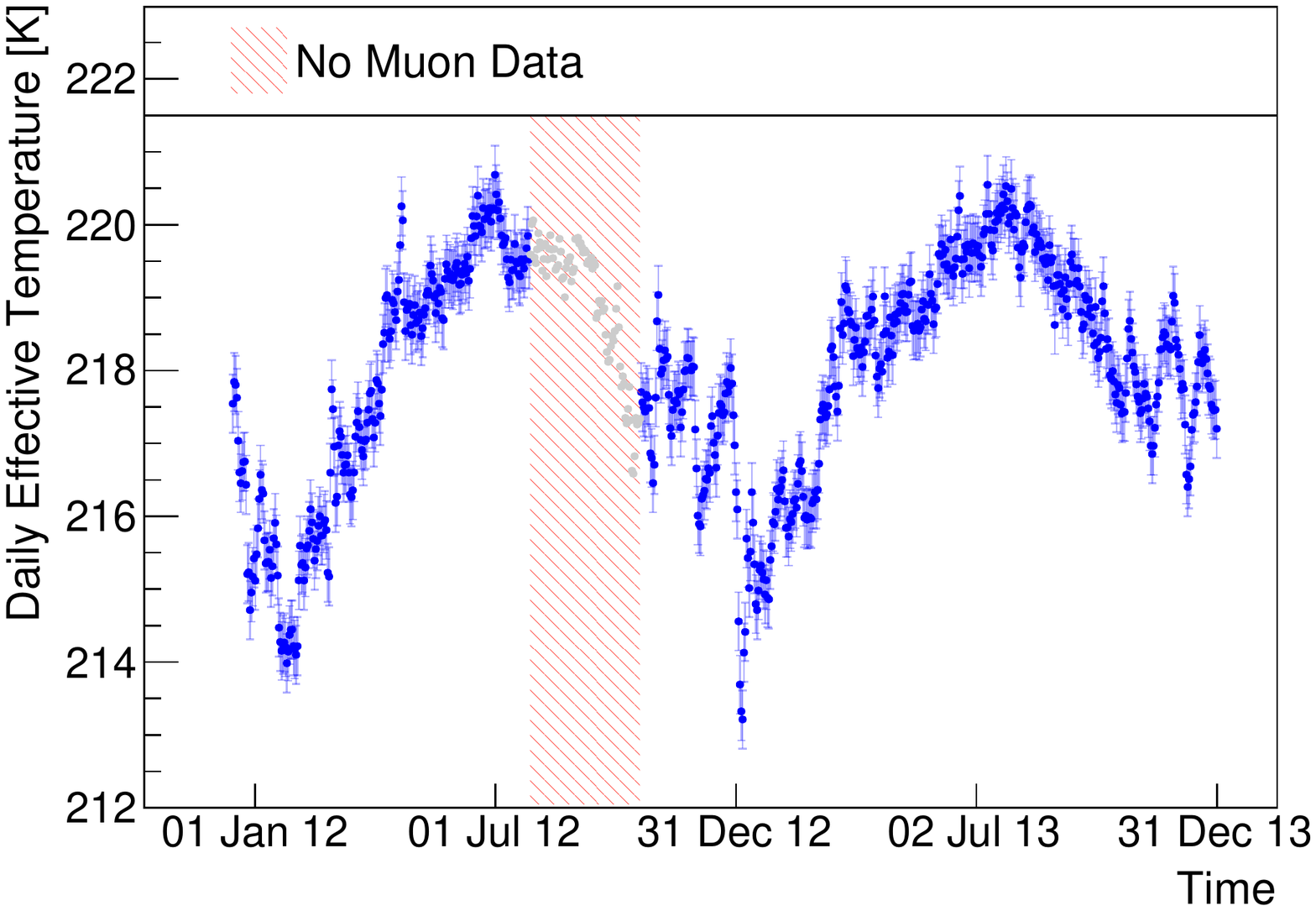}
\caption{\label{fig:temperature}(Left)  EH-dependent weights involved in the $\teff$ computation and time-averaged
atmospheric temperature, both as a function of pressure and altitude. (Right) Daily $\teff$ values computed using EH1 weights. ``No Muon Data'' refers to the 2012 summer shutdown.}
\end{figure}

Our goal is to use the atmospheric temperature to assess if and how it affects the muon production. However,
it is not possible to know at what altitude a cosmic muon is produced, hence
we follow~\cite{Barrett, Macro, Grashorn} and approximate the atmosphere with an
isothermal body characterised by an effective temperature $\teff$.
$\teff$ is defined as the temperature that would cause the observed muon intensity if the
atmosphere were isothermal~\cite{Barrett}, and it is computed as a weighted average of the temperature $T$ over the atmospheric depth $X$, with weights $W$, as shown in Eq.~\ref{eq:teff_int}.
\begin{equation}
\teff  = \frac{\int_{0}^{\infty} \mathrm{d}X \, T(X) \, W(X)}
{\int_{0}^{\infty} \mathrm{d}X \, W(X)} \simeq
\frac{\sum_{i} \Delta X_i \, T(X_i) \, W(X_i)}
{\sum_{i} \Delta X_i \, W(X_i)}
\label{eq:teff_int}
\end{equation}
The atmospheric depth $X$ is expressed in g/cm$^2$ and is related to the pressure level
by the relation 1~hPa = 1.019~ g/cm$^2$.
The approximation with the discrete summation is appropriate because the temperature data is available only at discrete pressure levels.

The weight $W$ associated with each pressure level reflects the model
 that we use to explain the modulation of the muon flux (see below).
The left panel of Fig.~\ref{fig:temperature} shows that pressure levels near the 
top of the atmosphere are weighted more heavily than the pressure levels at lower altitude.
The two main arguments to support this choice are:
(\textsc{i}) a high energy parent meson has less chance to decay after it reaches the high density
regions at low altitude, and (\textsc{ii}) very few of the parent mesons survive both nuclear interaction
and decay long enough to reach low altitude.

Both kaon and pion production and decay should be considered in the model~\cite{Barrett}.
However, because of limited sensitivity to the kaon contribution, older experiments computed their
weights using a pion-only model (e.g. Barrett~\cite{Barrett}, Sherman~\cite{Sherman},
Utah~\cite{Utah},  \textsc{Macro}~\cite{Macro}). This approach changed with MINOS~\cite{MinosFD} which,
for the first time, used a model put forward by Grashorn et al.~\cite{Grashorn} explicitly including
kaons.
Here we follow the latter approach, and we write $W(X_i) = W^{\pi}(X_i) + W^{K}(X_i)$, where
\begin{subequations}
\begin{equation}
W^{\pi,K}(X) \simeq \frac{\left(1-\frac{X}{\lambda_{\pi,K}}\right)^2 \,
e^{-\frac{X}{\Lambda_{\pi,K}}} \, A_{\pi,K}}
{\gamma + (\gamma+1) B_{\pi,K} \, K(X)
\left(\frac{\langle E_{\mathrm{thr}} \cos \theta \rangle }{\epsilon_{\pi,K}}\right)^2}
\end{equation}
\begin{equation}
K(X) \equiv \frac{X \, \left(1-\frac{X}{\lambda_{\pi,K}}\right)^2}
{\left(1-e^{-\frac{X}{\lambda_{\pi,K}}}\right) \, \lambda_{\pi,K}}
\end{equation}
\begin{equation}
\frac{1}{\lambda_{\pi,K}} = \frac{1}{\Lambda_N} - \frac{1}{\Lambda_{\pi,K}} \, \, .
\end{equation}
\label{eq:weights}
\end{subequations}

\noindent $A_{\pi ,K}$ is a constant comprising the amount of inclusive meson production in the forward fragmentation region, the masses of mesons and muons, and the muon spectral index. The parameter $B_{\pi ,K}$ accounts for the relative atmospheric attenuation of mesons. The parameter $\Lambda _{N,\pi ,K}$ is the atmospheric attenuation length of the cosmic ray primaries, pions and kaons, respectively. 
The meson critical energy, $\epsilon_{\pi,K}$, is the meson energy for which decay and interaction have an equal probability.
Finally, the parameter $\gamma$ is the muon spectral index.

The values of all the parameters are listed in Table~\ref{tab:weight_params} and are
inherited from \cite{MinosFD}, with the exception of
$\langle E_{\mathrm{thr}}\cos \theta \rangle$
calculated as described.
The left panel of Fig.~\ref{fig:temperature} shows three weight functions ---due to the fact that the $\ethr$ value
is site dependent---
together with the time-averaged temperature at the Daya Bay site. The $\teff$ daily values computed at
EH1 are shown in right panel of Fig.~\ref{fig:temperature}.

\begin{table}
\centering
\begin{tabular}{lrc}
  Parameter       & Value  & Reference\\
\hline
  $A_{\pi}$       & 1                      & \cite{MinosFD}         \\
  $A_K$           & $0.38 \cdot r_{K/\pi}$ & \cite{MinosFD}         \\
  $r_{K/\pi}$     & $0.149 \pm 0.06$       & \cite{Gaisser}$\dagger$ , \cite{Barr}$\ddagger$        \\
  $B_{\pi}$       & $1.460 \pm 0.007$      & \cite{MinosFD}         \\
  $B_{K}$         & $1.740 \pm 0.028$      & \cite{MinosFD}         \\
  $\Lambda_{N}$   & 120 \, g/cm$^2$        & \cite{Gaisser}         \\
  $\Lambda_{\pi}$ & 180 \, g/cm$^2$        & \cite{Gaisser}         \\
  $\Lambda_{K}$   & 160 \, g/cm$^2$        & \cite{Gaisser}         \\
  $\langle E_{\mathrm{thr}}\cos \theta \rangle_{\mathrm{EH1}}$ & $ 37  \pm 3 \,\mathrm{GeV} $  \\
  $\langle E_{\mathrm{thr}}\cos \theta \rangle_{\mathrm{EH2}}$ & $ 41  \pm 3 \,\mathrm{GeV}$  \\
  $\langle E_{\mathrm{thr}}\cos \theta \rangle_{\mathrm{EH3}}$ & $ 143 \pm 10 \,\mathrm{GeV}$  \\
  $\gamma$        & $1.7 \pm 0.1 $         & \cite{MinosChargeRatio}         \\
  $\epsilon_{\pi}$& $114 \pm 3 \, \mathrm{GeV} $ &  \cite{MinosFD}         \\
  $\epsilon_{K}$  & $851 \pm 14 \, \mathrm{GeV}$  & \cite{MinosFD}         \\
\end{tabular}
\footnotetext[1]{This reference is for the mean value only.}
\footnotetext[2]{This reference is for the uncertainty only.}
\caption{\label{tab:weight_params}Central values and uncertainties of the parameters used in Eq. \ref{eq:weights}. $\dagger$ Reference for the central value. $\ddagger$ Reference for the uncertainty.}
\end{table}

\subsection{Temperature Uncertainty}

To assess the uncertainty on the daily effective temperature $\sigma(\teff)$, we compute  $\teff$ again starting from a different temperature dataset, and we consider the spread of the difference between the new and old values to be a conservative estimate of the $\teff$ uncertainty.
The new dataset is the Integrated Global Radiosonde Archive (IGRA)~\cite{Durre} provided by the US National Climatic Data Center. It comprises temperature data from sounding balloons launched from many meteorological stations around the world, where the closest station to our detectors is located in the city of Shantou, Guandong (China), $235\,\mathrm{km}$ North-East of Daya Bay.
As mentioned in the previous section, the ECMWF dataset can be interpolated to an arbitrary location, hence --for the sake of this comparison-- we compute an ECMWF-based $\teff$ at the  Shantou's coordinates ($23^{\circ}21'$N, $116^{\circ}40'$E), and we compare it with the IGRA-based $\teff$. The distribution of the differences shows a spread of 0.4~K, which we propagate to the $\teff$ uncertainty.
As a further check, we also compare ECMWF-based $\teff$ values computed at Daya Bay's coordinates  with IGRA-based $\teff$ values computed at Shantou, and we find the spread to be the same.

The uncertainties associated with all the parameters involved in Eq.~\ref{eq:weights}
also contribute an additional temperature systematic uncertainty. To evaluate the impact, we proceed as follows:
(\textsc{i}) we associate to each parameter
a Gaussian distributed independent random variable, where the mean value and the sigma are chosen
according to the values in Table~\ref{tab:weight_params}, (\textsc{ii}) we generate $10^5$ weight functions
(like those shown in Fig.~\ref{fig:temperature}) using a random point in the parameter phase-space,
(\textsc{iii}) with each generated weight function we carry out a new $\teff$ calculation,
resulting in $10^5$ smeared $\teff$ values per day.
We build a daily distribution of the difference between the smeared and the nominal $\teff$ values,
and we find that over the whole data-taking period the spread induced by smearing the weights is
$0.15\,$K, $0.15\,$K, $0.07\,$K, respectively for EH1, EH2 and EH3. Combining these
uncertainties with the $0.4\,$K $\teff$ uncertainty common to all the experimental halls, we get
a total $\sigma ( \teff ) $ of $0.43\,$K,  $0.43\,$K and  $0.41\,$K respectively.

%
%
\section{Correlation Analysis}

\begin{figure}[H]
\centering
\begin{tabular}{cc}
\includegraphics[width=0.4\textwidth]{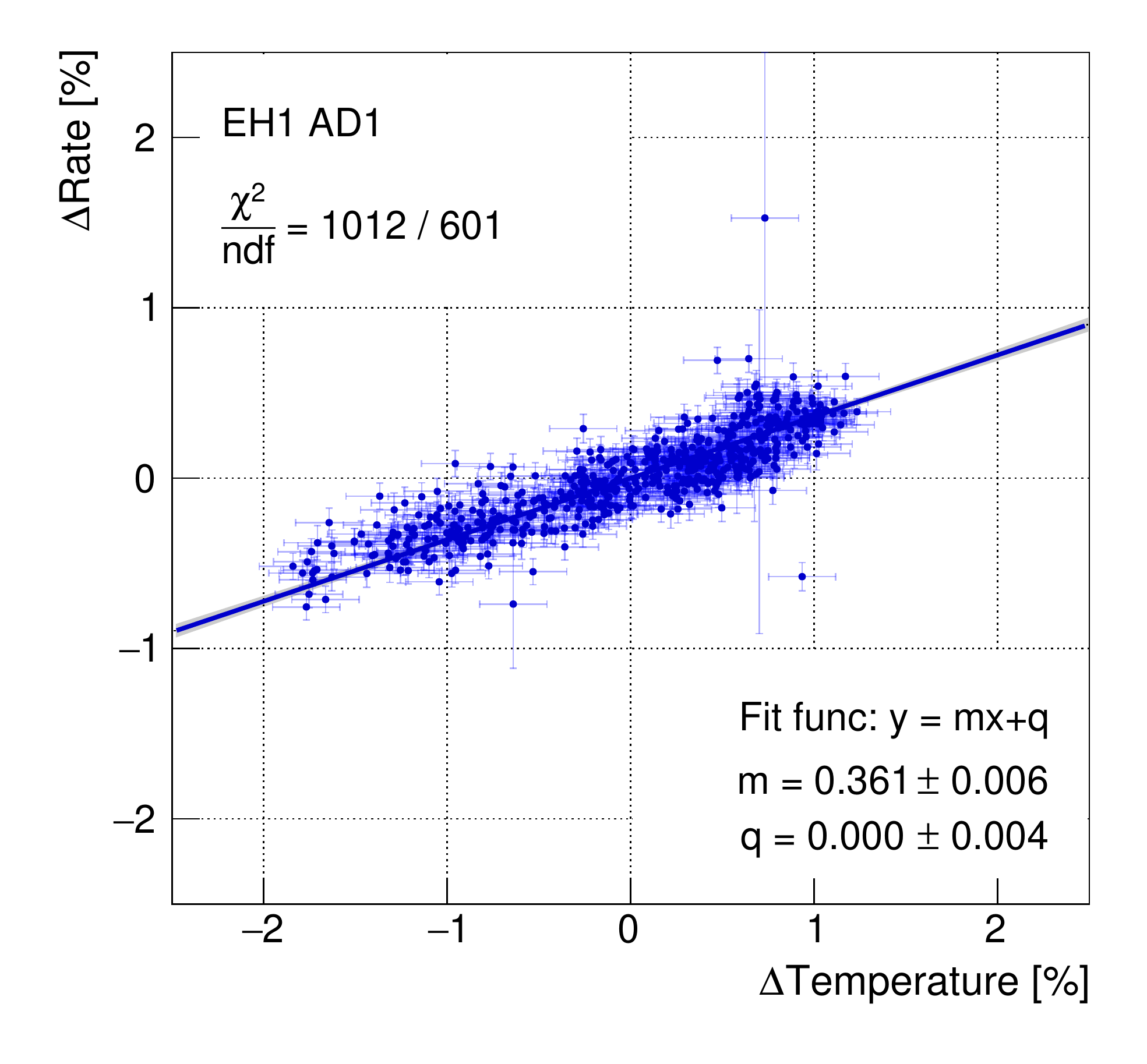} &
\includegraphics[width=0.4\textwidth]{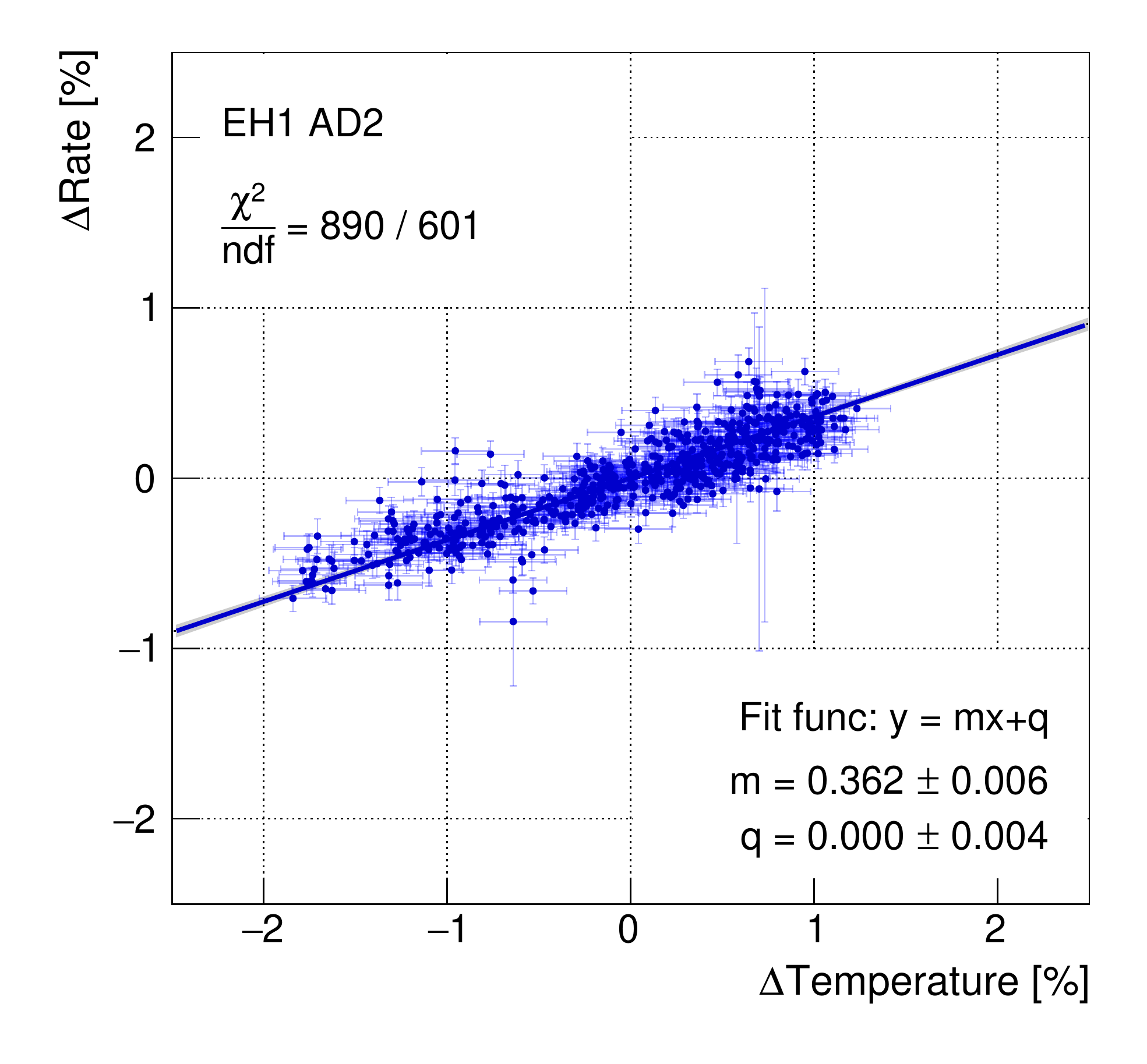}\\

\includegraphics[width=0.4\textwidth]{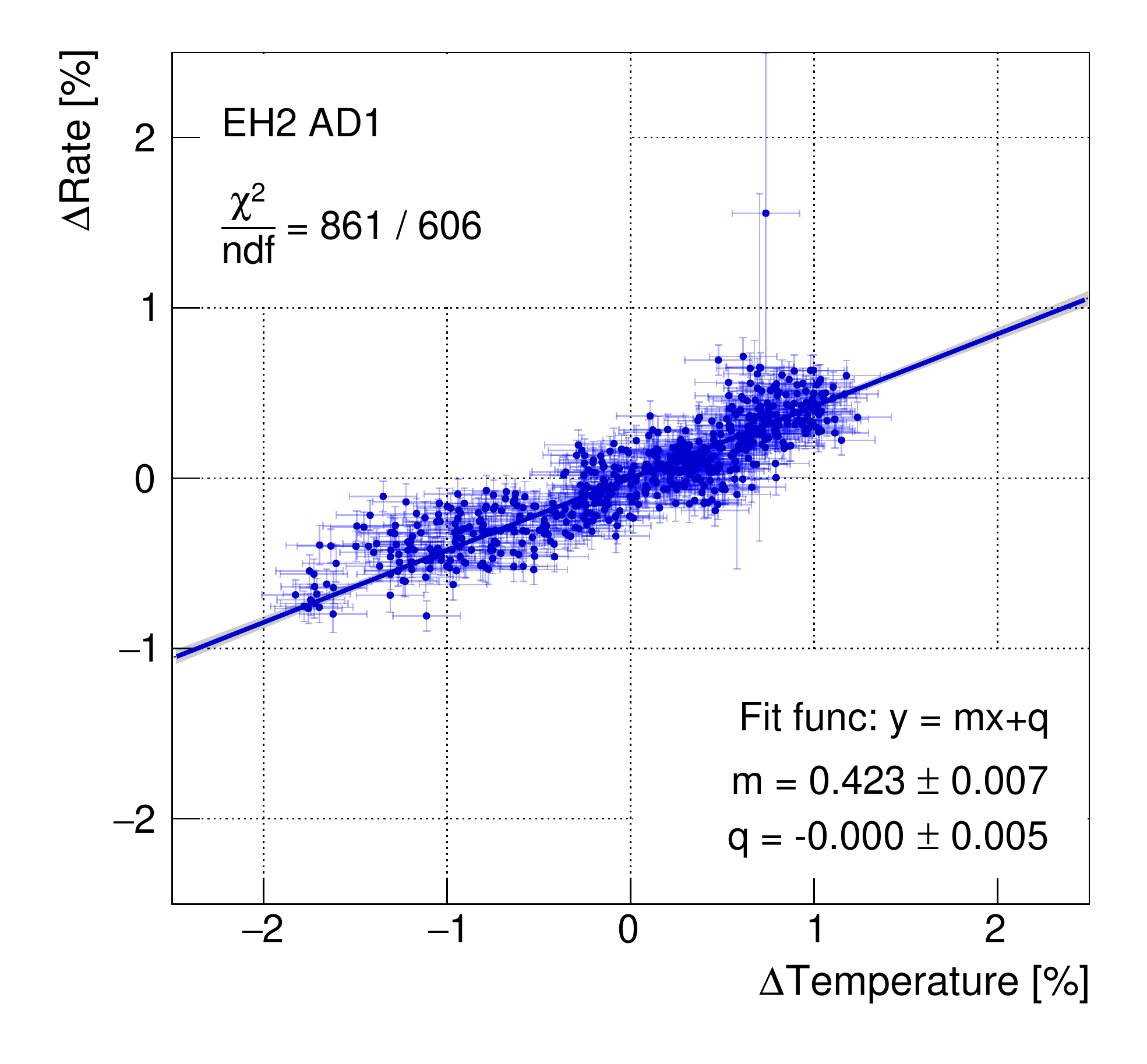}&
\includegraphics[width=0.4\textwidth]{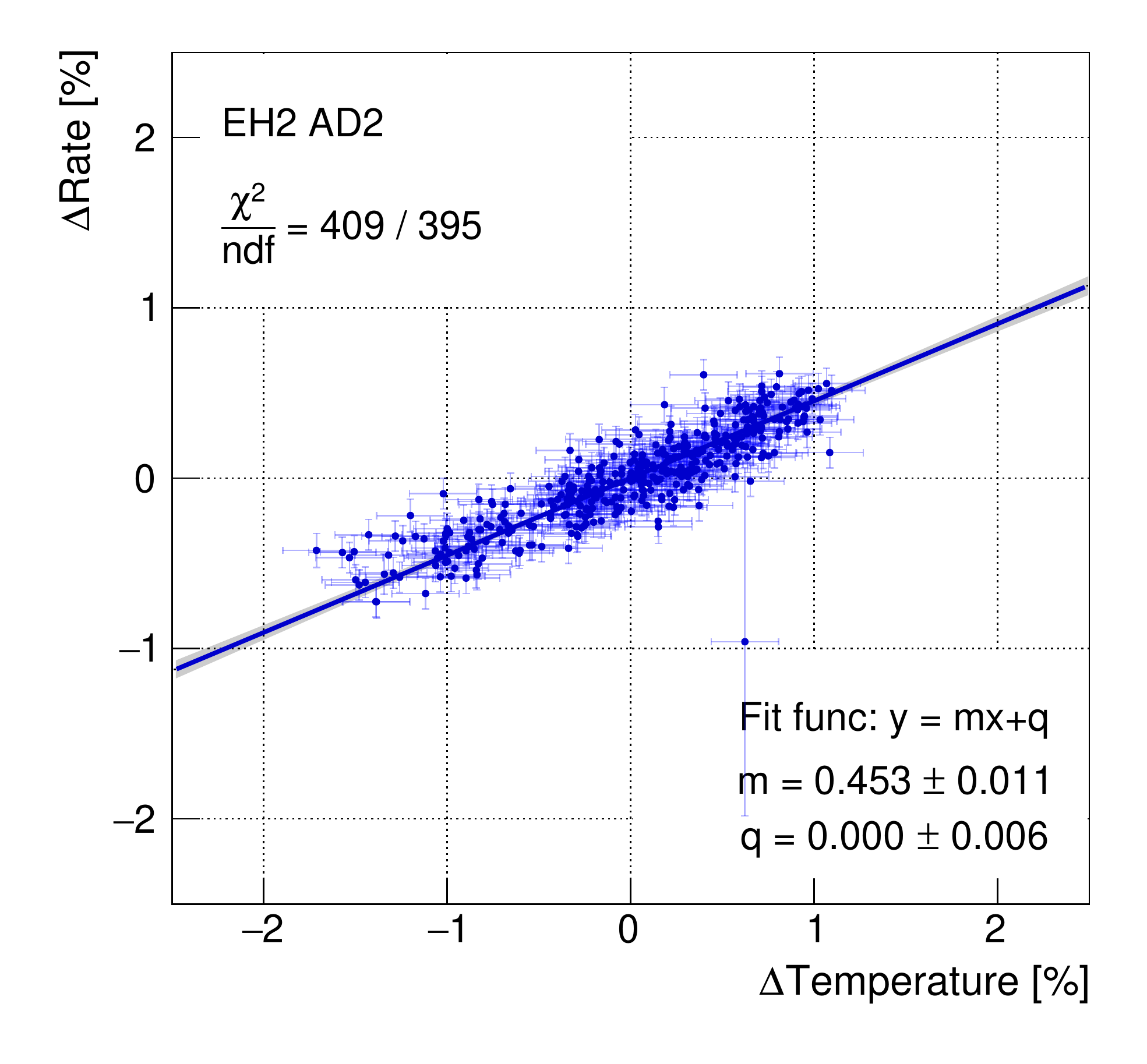}\\

\includegraphics[width=0.4\textwidth]{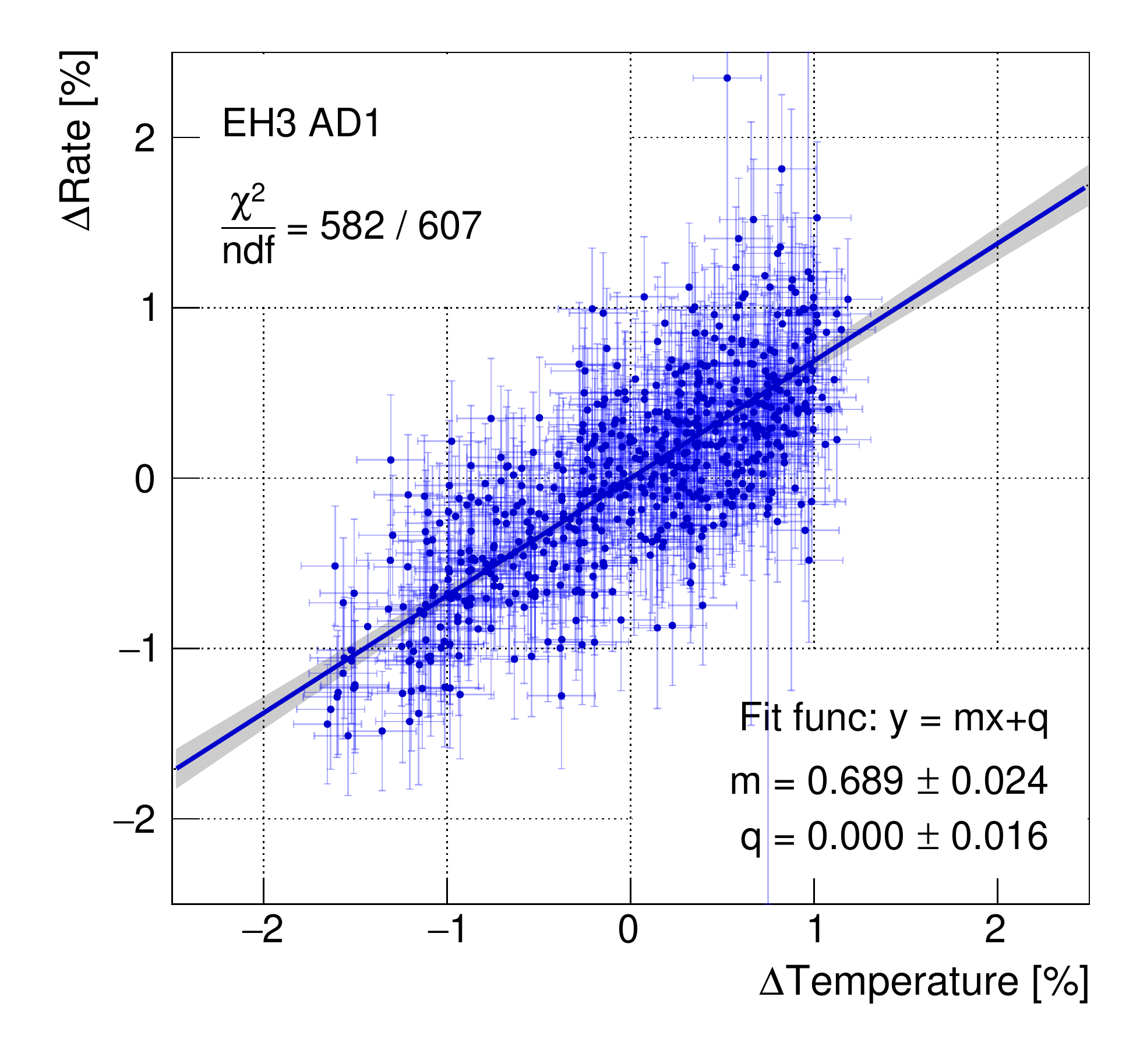} &
\includegraphics[width=0.4\textwidth]{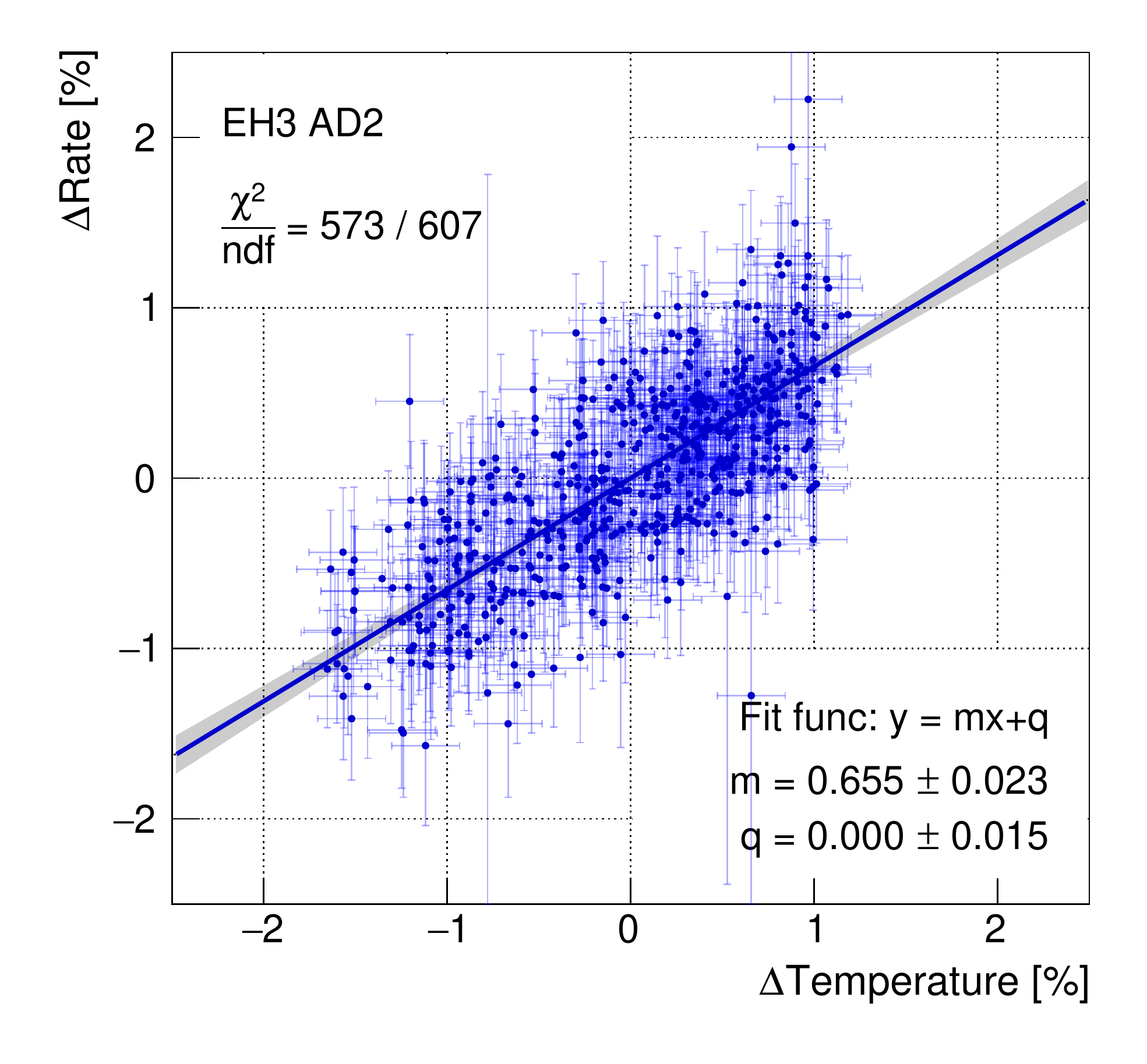}\\

\includegraphics[width=0.4\textwidth]{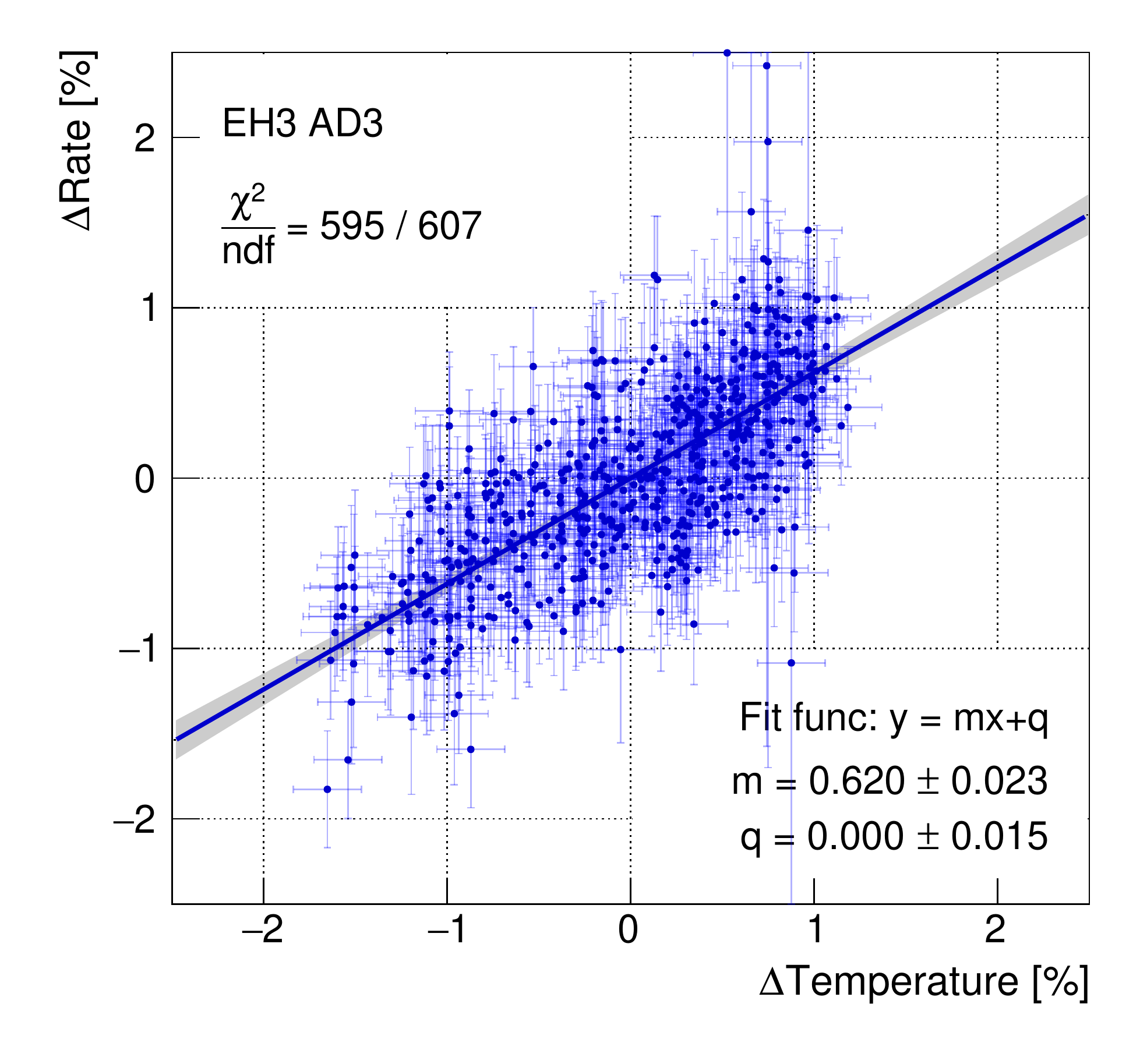}&
\includegraphics[width=0.4\textwidth]{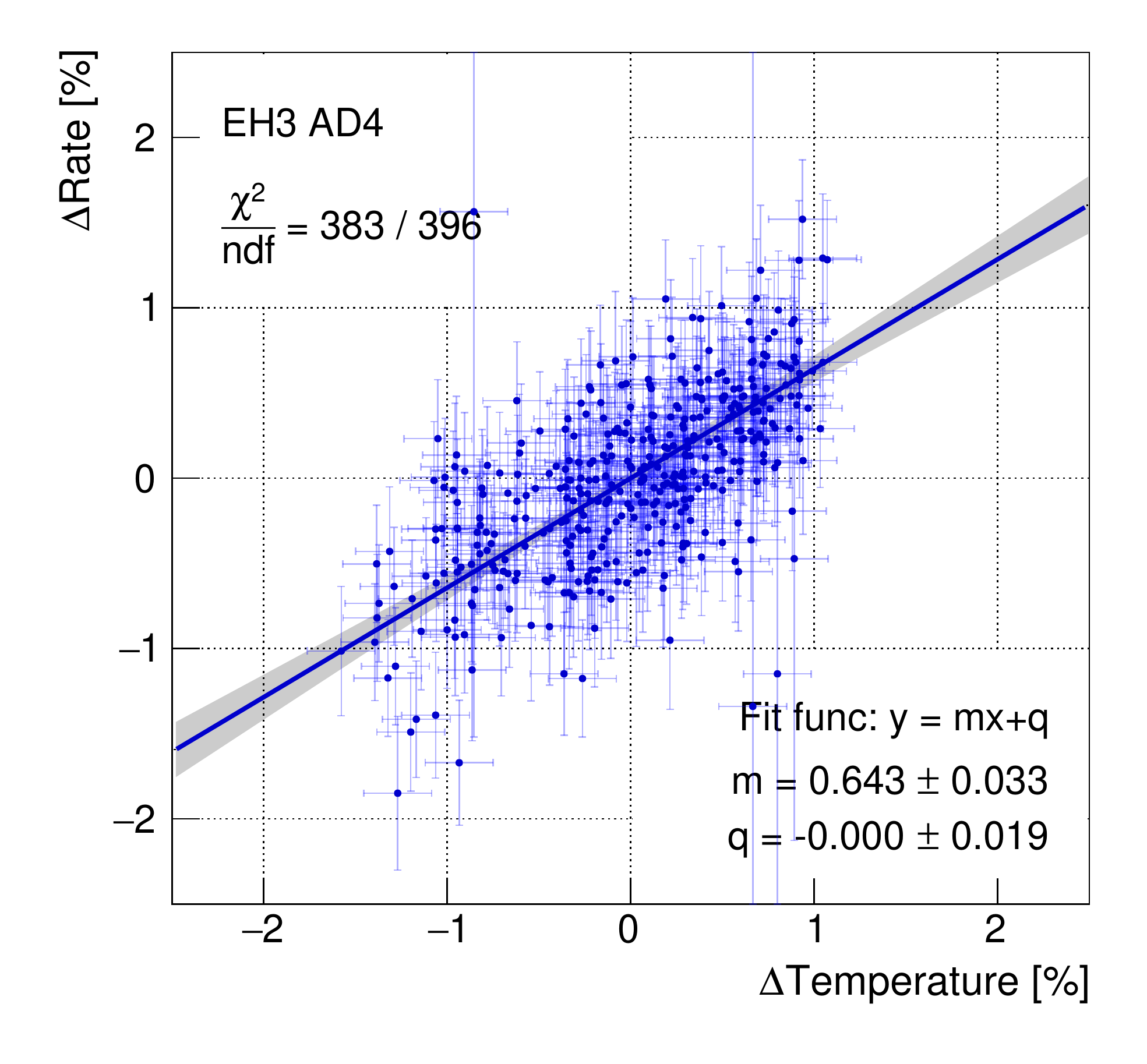}\\
\end{tabular}
\caption{Relative muon rate variation vs relative effective temperature variation as measured in the
eight ADs, together with the result of a linear regression accounting for uncertainties on both variables.}
\label{fig:correlation_per_ad}
\end{figure}

The aim of this analysis is to assess quantitatively how a variation in the atmospheric temperature
relates to a variation in the underground muon rate. For each AD we start from a daily binned $\teff$
dataset (common to all the ADs within an experimental hall) and from a daily binned muon rate dataset,
and we build a scatter plot where the x (y) axis represents the temperature (muon rate) relative
variation with respect to its mean value. All the scatter plots are shown in Fig.~\ref{fig:correlation_per_ad}.
The y error bar on each data point represents the Poissonian uncertainty on the number
of detected muons, while the x error bar represents the temperature uncertainty obtained by comparing
two temperature datasets (see previous section).

A linear regression accounting for errors on both axes is performed to each scatter plot. We use the
fitting routines provided by ROOT~\cite{root}, and we define the slope of the
fitted linear function to be the correlation coefficient $\alpha$.

\subsection{Systematics Study}

Systematic uncertainties affecting the correlation coefficients can be divided into regression-related and muon-related.
The former deal with the absence of two complete oscillation cycles in the muon data. 
No AD continuously acquires data for more than one period due to the Summer 2012 shutdown and the lack of 
the first seven months of data for the two ADs.
To check the impact of computing the correlation coefficient on a limited time range,
we use EH1 AD1, EH1 AD2 and EH2 AD1 data, since these ADs have reduced statistical uncertainties
(muon rates are higher because
the experimental sites are shallower), and they collected data both before and after the shutdown.
Using these ADs, we compute two separate correlation coefficients for the pre- and post-shutdown periods, and we find a
maximum discrepancy of 8\%, which we take to be the systematic uncertainty associated with the fitting range.

We define muon-related systematics to be those uncertainties that are expected not only to inflate the
correlation coefficient uncertainty, but also to bias its central value
(as opposed to the statistical uncertainties that are already included in the linear regression).
For this reason, muon-related systematics  are evaluated by performing new linear regressions with slightly different muon datasets.
The two effects we want to study are the AD-dependent correction of the energy drift over time, and the tighter energy cut for EH3 AD1, resulting from the liquid scintillator leakage into the MO region.
All the energy drift corrections are below $1\%$. As a consequence, we conservatively shift the energy thresholds used in the muon
selection criteria up and down by 1\%, and we perform new linear regressions based on the higher/lower energy datasets.
The corresponding correlation coefficients deviate at most 3\% from the nominal values, hence we assign a 3\% systematic
uncertainty  to all the ADs.
The effect of the tighter energy cut on EH3 AD1 is two-fold. (\textsc{i})
Daya Bay electronics distinguishes two energy ranges, ``fine'' and ``coarse'', depending on the PMT output charge. 
%
%
\noindent The ``fine'' (lower) range is well calibrated since it is used for precision neutrino physics, while the higher range is meant 
only to tag cosmic muons and is less precisely calibrated. The event-level energy threshold of $100\mev$ 
therefore results in some of the channels operating  in the ``coarse'' energy range, whose stability over time is more uncertain.
To assess the impact of these effects we raise the energy cut from
$60\mev$ to $100\mev$ in the EH1 and EH2 ADs, and we look at variation in the $\alpha$ values. The reason for considering
EH1 and EH2 is to exploit their larger muon flux and to avoid that a shift in $\alpha$ might be ascribed to statistical
fluctuations. This test results in a $2\%$ variation of the $\alpha$ value.
(\textsc{ii}) The tighter energy cut might not efficiently reject all the muons passing through the MO region,
hence there might be a bias introduced by this residual contamination. To assess its relevance, we
compare the number of events with $R>2\,\mathrm{m}$ across all the EH3 ADs, and we see that, despite a higher energy cut,
EH3 AD1 collects more events. Moreover, the event excess is confined to post-Summer 2012, increasing our confidence
in ascribing those events to the LS leak. We further evaluate the outcome of such contamination on the correlation coefficient
by performing the EH3 AD1 regression on a new muon dataset comprising an additional $R<2\,\mathrm{m}$ vertex cut.
We find the difference between the new and the  nominal correlation coefficient to be at the level of 4\%. The total systematic
uncertainty affecting EH3 AD1 is therefore 5.4\%.

Results are summarised in Table~\ref{tab:corr_coeff}.

\subsection{Correlation Coefficient Combination}
\begin{figure}
\centering
\begin{tabular}{ccc}
\includegraphics[width=0.31\textwidth]{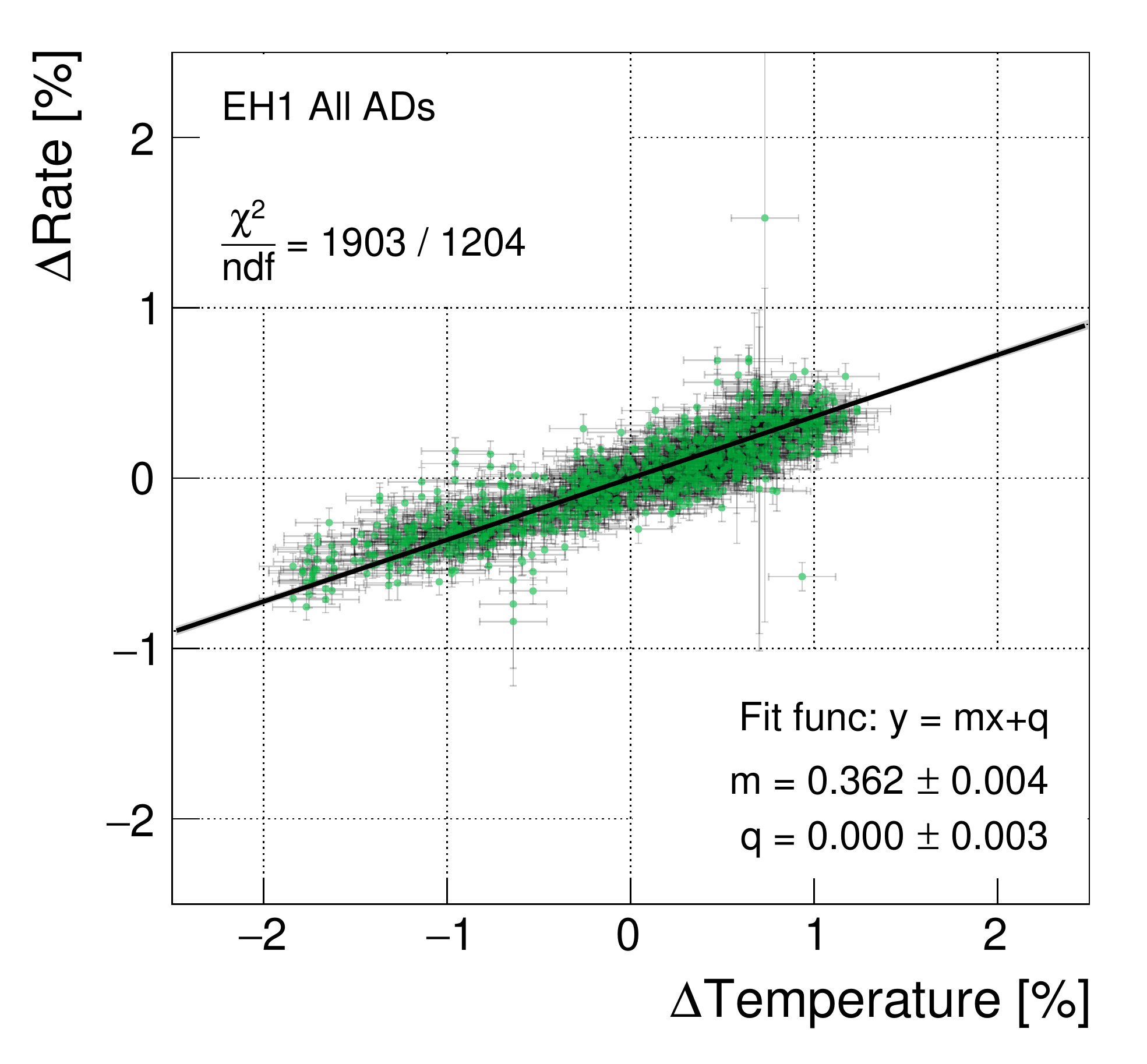} &
\includegraphics[width=0.31\textwidth]{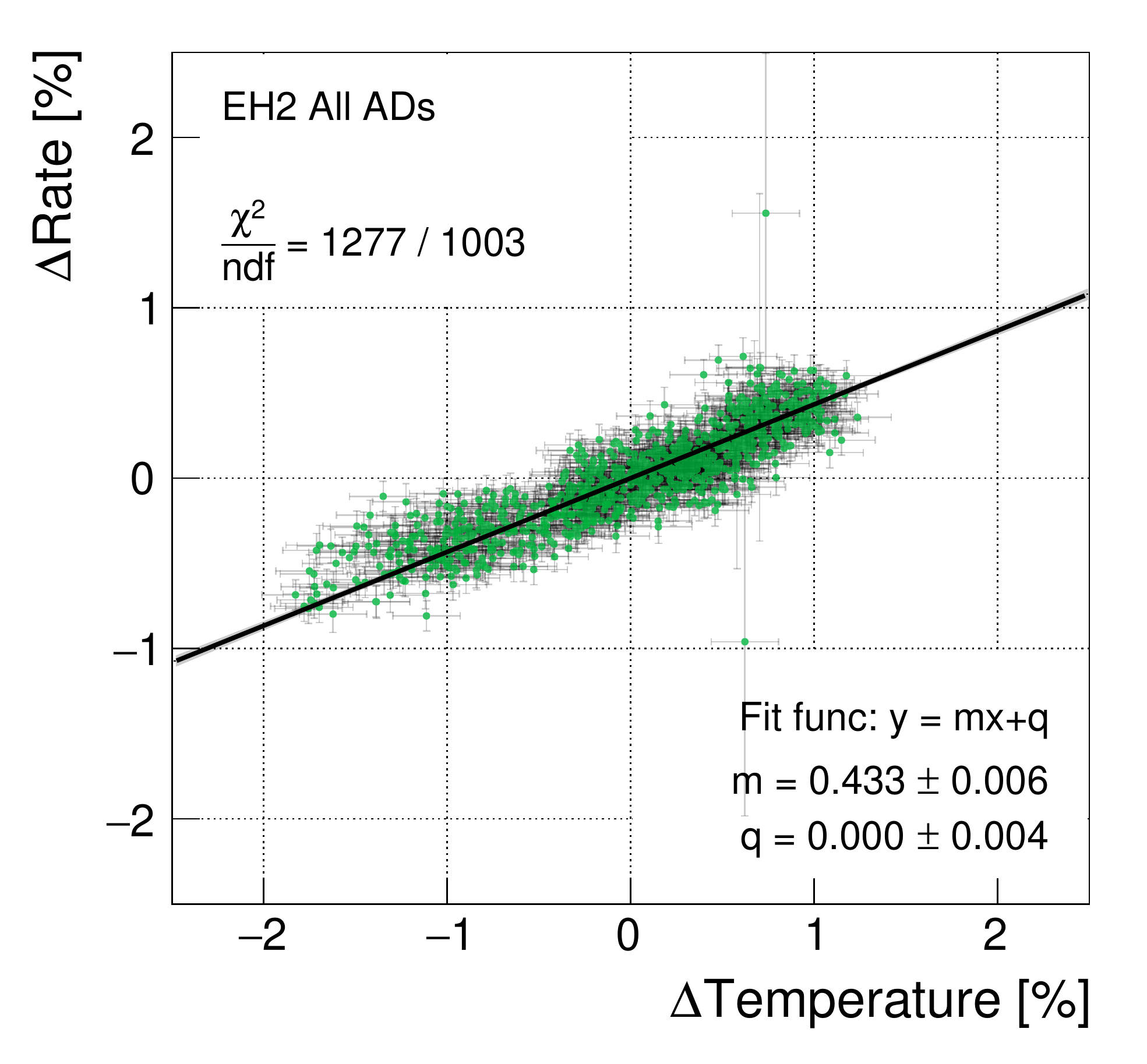} &
\includegraphics[width=0.31\textwidth]{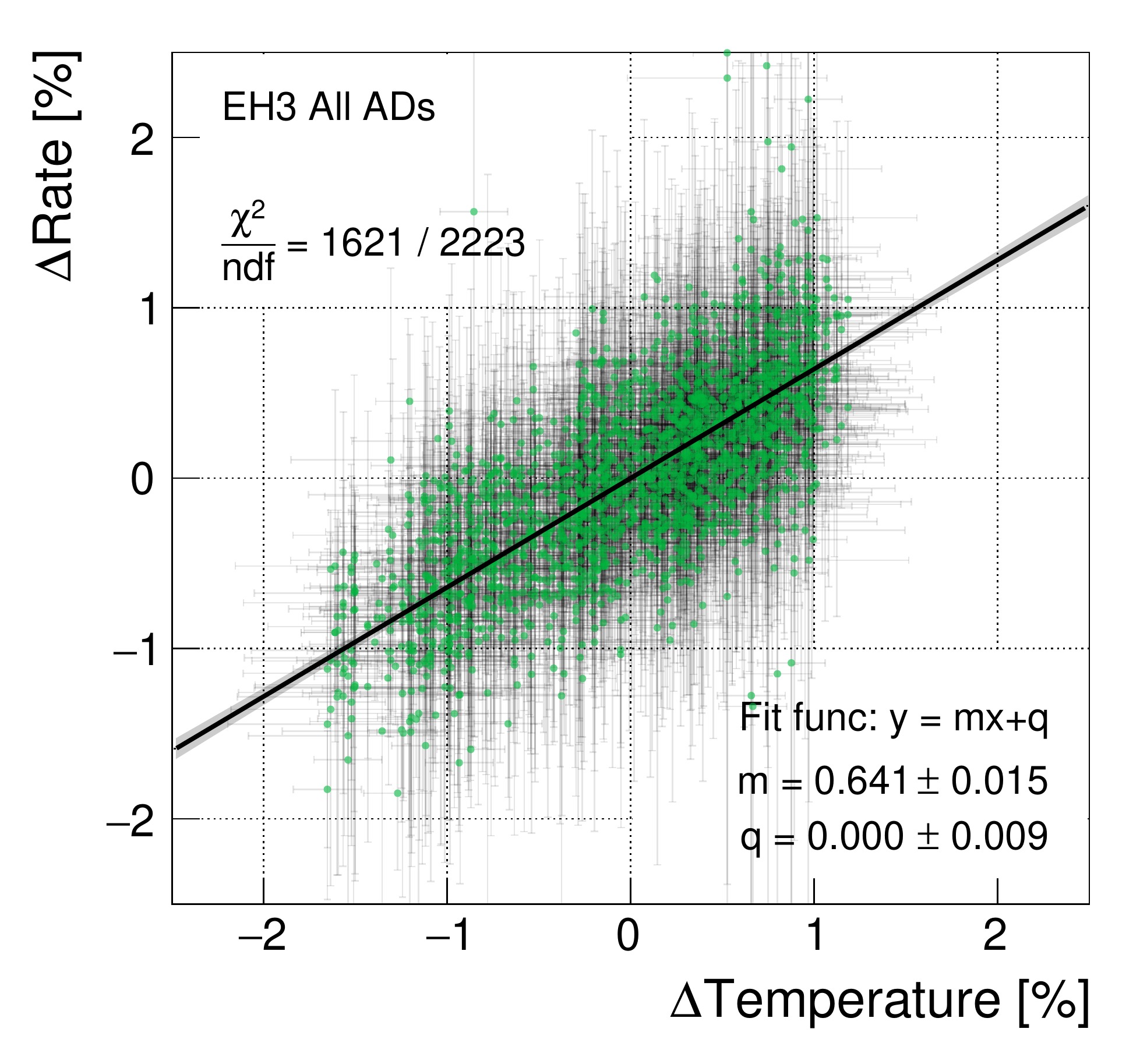}\\
\end{tabular}
\caption{Relative muon rate variation vs relative effective temperature variation constructed by merging data from ADs belonging to the same experimental hall, together with the result of a linear regression accounting for uncertainties on both variables}
\label{fig:scatter_combined}
\end{figure}

\begin{table}
\centering
\begin{tabular}{ccccccc}
\multicolumn{2}{c}{Detector} & Correlation & \multicolumn{3}{c}{Uncertainty} & Combined      \\
                           & & Coefficient & Stat. & Sys. & Tot. &   \\
\hline
\rule{0pt}{3ex}
\multirow{2}{*}{EH1} & AD1 & 0.3614 & 0.0062 & 0.0309 & 0.0315 & \multirow{2}{*}{$0.362 \pm 0.031$} \\ 
                     & AD2 & 0.3624 & 0.0062 & 0.0310 & 0.0316 & \\
\rule{0pt}{3ex}
\multirow{2}{*}{EH2} & AD1 & 0.4232 & 0.0071 & 0.0362 & 0.0369 & \multirow{2}{*}{$0.433 \pm 0.038$} \\ 
                     & AD2 & 0.4530 & 0.0107 & 0.0387 & 0.0402 & \\
\rule{0pt}{3ex}
\multirow{4}{*}{EH3} & AD1 & 0.6890 & 0.0240 & 0.0664 & 0.0707 & \multirow{4}{*}{$0.641 \pm 0.057$} \\ 
                     & AD2 & 0.6549 & 0.0234 & 0.0560 & 0.0607 & \\
                     & AD3 & 0.6200 & 0.0234 & 0.0530 & 0.0579 & \\
                     & AD4 & 0.6427 & 0.0334 & 0.0549 & 0.0643 & \\
\end{tabular}
\caption{\label{tab:corr_coeff}
Experimental values of the correlation coefficient ($\alpha$) per detector.}
\end{table}

The correlation coefficient between the muon rate variation and the atmospheric temperature variation is known to
increase as a function of the overburden, as a result of a harder muon energy spectrum. For this reason we 
combine the results obtained from ADs sharing the same experimental hall, with the aim to provide our results as a
function of $\langle \ethr \cos \theta \rangle$.
Coefficient uncertainties in the same experimental hall are partially correlated, since they share the same temperature dataset.
Instead of combining the values, we choose to merge the raw datasets and to perform a new linear regression on the
combined scatter plots, as shown in Fig.~\ref{fig:scatter_combined}.
This procedure has also the advantage that no fit relies on a truncated dataset --- such as EH2 AD2, and EH3 AD4.

Systematic uncertainties common to all the ADs can directly be applied to the combined correlation coefficients.
However special care must be devoted to handling EH3 AD1. Indeed this AD has its own systematic uncertainty
resulting from the LS leak, and must be weighted accordingly in the EH3 combination. To this end we exploit a feature
of the linear regression, namely the fact that applying a scale factor to the error bars of the data points being fitted, results
in  a proportional scaling of the fit uncertainty. We therefore scale the error bars of the EH3 AD1 data points
by the amount $\sigma_{\mathrm{tot}} / \sigma_{\mathrm{stat}}$, where
$\sigma_{\mathrm{tot}} = \sqrt{ \sigma^2_{\mathrm{stat}} + \sigma^2_{\mathrm{sys} }}$.
In this way, fitting the EH3 AD1 dataset with inflated uncertainties results in a fit uncertainty which includes the systematic one.
As a result, the EH3 AD1 data points are now correctly weighted in the combined EH3 fit.
The correlation coefficients of the three experimental halls are summarised in Table~\ref{tab:corr_coeff}.

%
%
\section{Comparison With Model Prediction and Other Experiments}

\begin{figure}
\centering
\includegraphics[width=0.8\linewidth]{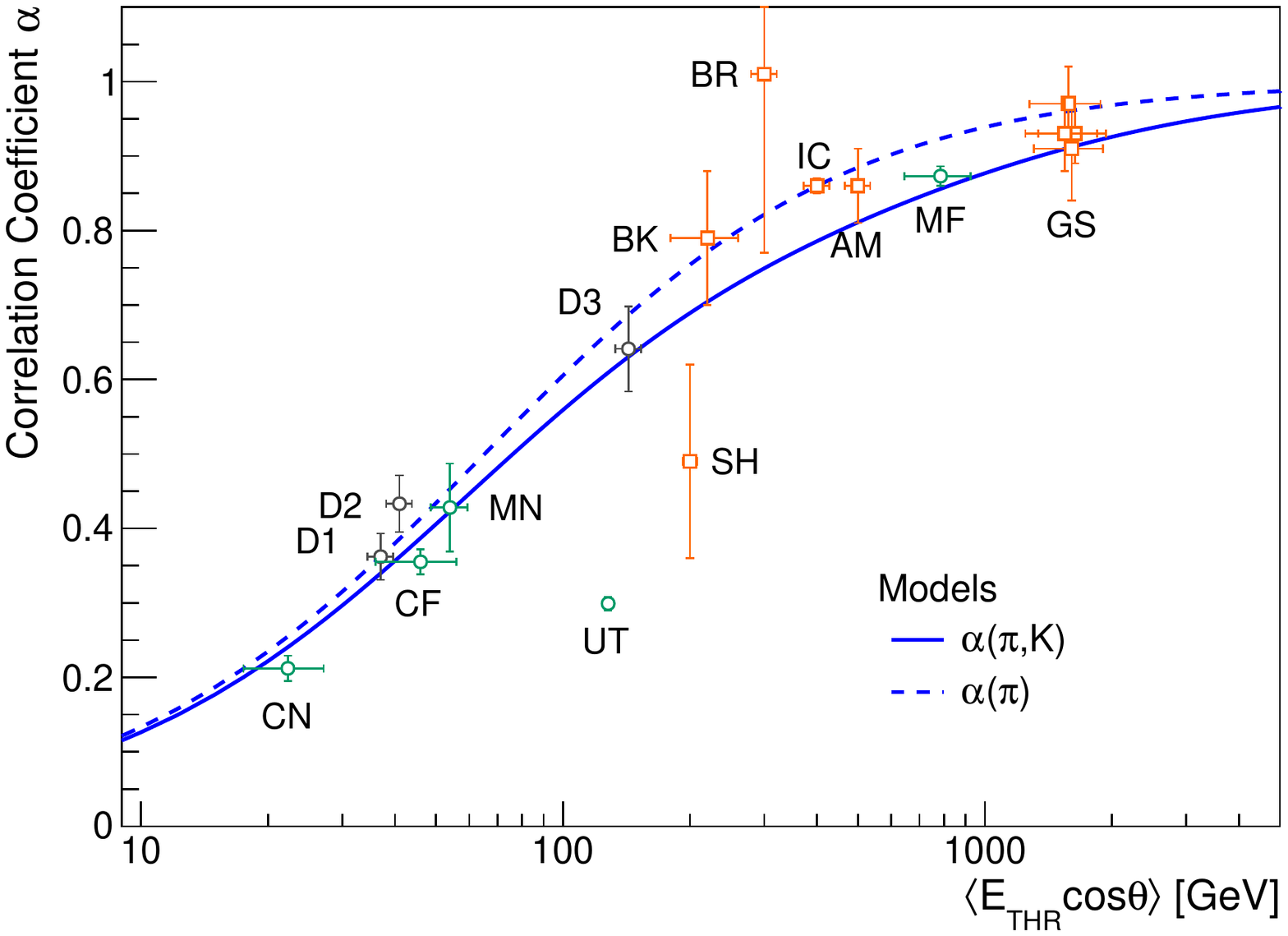}
\caption{\label{fig:comparison_other_exp}Comparison of the experimental $\alpha$ values with the  model accounting for pions
and kaons (solid line), and with the model accounting for pions only (dashed line). Values determined in this analysis are reported as
D1, D2 and D3 respectively for the three experimental halls. Other experiments include Amanda (AM)~\cite{Amanda}, Baksan (BK)~\cite{Baksan},
Barrett (BR)~\cite{Barrett2}, Icecube (IC)~\cite{Icecube2011}, MINOS Near (MN)~\cite{MinosND} and far (MF)~\cite{MinosFD} detector, 
Double Chooz Near (CN) and Far (CF) detectors~\cite{DC}, Sherman (SH)~\cite{Sherman}, and Utah (UT)~\cite{Utah}. 
The four Gran Sasso (GS) based measurements are  \textsc{Macro}~\cite{Macro}, Borexino~\cite{Borexino},
and the two Gerda values~\cite{Gerda}. Their $\langle \ethr  \cos \theta \rangle$ are artifically displaced
on  the horizontal axis for the sake of visualization.
The values and uncertainties of $\langle \ethr  \cos \theta \rangle$ shown in the Figure for experiments
tagged with a square marker have been estimated as described in the text.}
\end{figure}

The model prediction for $\alpha$  can be written as:
\begin{equation}
\label{eq:theoretical_alpha1}
\alpha = \frac{\teff}{I_{\mu}} \, \frac{\partial I_{\mu}}{\partial \teff} \, .
\end{equation}
Grashorn et al.~\cite{Grashorn} show that this prediction can be expressed in terms of both pion and kaon contributions as
\begin{equation}
\label{eq:theoretical_alpha2}
\alpha = \frac{1}{D_{\pi}} \frac{1/\epsilon_K + A_K (D_{\pi} / D_{K})^2 / \epsilon_{\pi}}{1/\epsilon_{K} + A_K (D_{\pi}/D_{K}) / \epsilon_{\pi}} \, \, ,
\end{equation}
where
\begin{equation}
D_{\pi ,K} = \frac{\gamma}{\gamma + 1} \, \frac{\epsilon_{\pi ,K}}{1.1 \,  \langle \ethr \cos \theta \rangle} + 1 \, .
\label{eq:ethr_dependency}   
\end{equation}
Equation~\ref{eq:theoretical_alpha2} can be reduced to \textsc{Macro}'s previously published $\alpha$~\cite{Macro}, which is only valid for pion-induced muons, by setting $A_K = 0$ (i.e. no kaon contribution).
To get the model prediction of $\alpha$ for the three Daya Bay sites, we plug the parameter values listed in Table~\ref{tab:weight_params} into Eq.~\ref{eq:theoretical_alpha2}. We consider both the model accounting for $\pi$ and K and the model accounting for $\pi$ only, and we report our findings in
Table~\ref{tab:corr_coeff_prediction}. The two models are shown in Fig.~\ref{fig:comparison_other_exp}.

To quantify the systematic uncertainty associated with the model prediction, we smear each input parameter according to its uncertainty, and in Table~\ref{tab:sys_unc_on_prediction} we assess its impact on the $\alpha$ prediction. As expected, the $\langle  \ethr \cos \theta \rangle$ uncertainty is driving the overall error budget.
By comparing the experimental with the predicted
 $\alpha$ values (Table~\ref{tab:corr_coeff_prediction}), it can be noticed that the former are consistently larger than the latter, hence favoring a lower kaon contribution with respect to the one currently used in literature ($r_{K/\pi}=0.149$~\cite{Gaisser, MinosFD, Borexino, MinosND}).

Figure~\ref{fig:comparison_other_exp} shows how the Daya Bay result compares to other
experiments\footnote{Experiments such as Torino~\cite{Castagnoli}, Hobart~\cite{Fenton},
Poatina~\cite{Poatina1, Poatina2}, and
Matsushiro~\cite{Matsushiro} are not included in this comparison, because
they perform a combined regression trying to correlate their muon rate with the pressure measured at sea
level (barometric coefficient) and the atmospheric temperature (temperature coefficient) at the same time.
Such measurements cannot be easily compared in a plot with all the others, which are based
only on the correlation between muon rate and atmospheric temperature.}
and to the model prediction. All the correlation coefficients are presented as a function of
$\langle \ethr  \cos \theta \rangle$, since this is the only free parameter in Eq.~\ref{eq:ethr_dependency}.
We stress here that such a quantity is sensitive not only to the vertical depth, but also to the shape of the overburden.
Indeed, the muon energy loss is determined by the amount of material crossed.
In the case of a flat overburden, however, $\langle \ethr  \cos \theta \rangle$ can be well approximated with the
minimum value of $\ethr$, namely the value evaluated along the vertical direction $\ethr(\theta =0)$. For this reason many experiments
quote only the $\ethr$ quantity. To assess to what extent this approximation holds, we use a toy Monte Carlo to scan
the quantity $\langle \ethr(\theta)  \cos \theta \rangle / \ethr(0)$ at various depths, and we find that the deviation grows
monotonically with the overburden, and stays below 10\% up to $1.6\,\mathrm{kmwe}$.
For experiments not quoting any uncertainty on $\ethr$, we use this deviation to associate a horizontal error bar
to the points shown in Fig.~\ref{fig:comparison_other_exp}. Two exceptions are the Baksan and Gran Sasso experiments.
The Baksan overburden is non-flat, and the average cosmic muon zenith angle is $\langle\theta \rangle = 35^{\circ}$~\cite{Baksan2}.
We consider  $\ethr \cos \langle \theta \rangle$ to be a lower bound to  $\langle \ethr  \cos \theta \rangle$, hence we use the
Baksan's quoted $\ethr$ as the central value, and the difference between $\ethr$ and  $\ethr \cos \langle \theta \rangle$  as the
uncertainty. The \textsc{Macro} collaboration measured the minimum $\ethr$ at the Gran Sasso laboratory to be 1300 GeV.
From their  $\alpha$ prediction based on a Monte Carlo simulation accounting for the Gran Sasso mountain
profile, it can be inferred that $\langle \ethr  \cos \theta \rangle \sim 1600 \, \mathrm{GeV}$. For all Gran Sasso experiments
we take the latter as central value, and the difference with respect to $\ethr$ as uncertainty.
Such an uncertainty is large enough to accomodate Borexino's and Gerda's $\ethr$ values, which are based on Grashorn's flat approximation~\cite{Grashorn}.

\begin{table}
\centering
\begin{tabular}{cccc}
Exp. Hall & \multicolumn{2}{c}{Prediction} & This Work \\
\cline{2-3}
 & Including K and $\mathrm{\pi}$ \cite{Grashorn} & Including $\pi$ only \cite{Macro} &\\
\hline
EH1 & $0.340 \pm 0.019$ & $0.362 \pm 0.018$ & $0.362 \pm 0.031$ \\
EH2 & $0.362 \pm 0.019$ & $0.386 \pm 0.018$ & $0.433 \pm 0.038$ \\
EH3 & $0.630 \pm 0.019$ & $0.687 \pm 0.018$ & $0.641 \pm 0.057$ \\
\end{tabular}
\caption{\label{tab:corr_coeff_prediction}
Predicted and measured values of the correlation coefficient at the different experimental halls.}
\end{table}

\begin{table}
\centering
\begin{tabular}{ccS[table-format=1.2]}
\multicolumn{2}{c}{Systematic Uncertainty} & \multicolumn{1}{c}{Resulting} \\
\cline{1-2}
\multicolumn{1}{c}{Parameter} & \multicolumn{1}{c}{Magnitude} & \multicolumn{1}{c}{Uncertainty on $\alpha$} \\
\hline
 $\langle \ethr \cos\theta \rangle$  & $7.0\%$ & 0.015       \\
 $R_{K / \pi}$    & $4.0\%$ & 0.0085      \\
 $\epsilon_{\pi}$ & $2.6\%$ & 0.0055      \\
 $\epsilon_{K}$   & $1.6\%$ & 0.00011     \\
 $\gamma$         & $5.9\%$ & 0.0049     \\
\hline
Total & & 0.019 \\
\end{tabular}
\caption{\label{tab:sys_unc_on_prediction}Systematic uncertainties affecting the theoretical prediction of the correlation coefficient.}
\end{table}

%
%
\section {Conclusion}
Muon rate variations over a time period of two years were measured using all the active components of
the Daya Bay experiment. Muon rates were then correlated with the effective atmospheric temperature
above the three experimental halls. The effective temperature is the result of a weighted average over
all the pressure levels comprised in the raw temperature dataset provided by the European Centre for
Medium-Range Weather Forecasts, and is needed to account for the fact that the majority of the muons
are produced in the uppermost layers of the atmosphere.
The correlation coefficient $\alpha$ at the three experimental halls was found to be:

\begin{center}
\begin{tabular}{lcl}
$ \alpha_{\text{EH1}} = 0.362\pm0.031$ & at  & $\langle \ethr \cos \theta \rangle_{\text{EH1}} = 37 \pm 3$ GeV\\
$ \alpha_{\text{EH2}} = 0.433\pm0.038$ &  at &  $\langle \ethr \cos \theta \rangle_{\text{EH2}} = 41 \pm 3$ GeV\\
$ \alpha_{\text{EH3}} = 0.641\pm0.057$ &  at &  $\langle \ethr \cos \theta \rangle_{\text{EH3}} = 143 \pm10$ GeV
\end{tabular}
\end{center}

The importance of this measurement lies in the fact that Daya Bay is
able to probe the muon seasonal variation at different overburdens using identically designed detectors.
Moreover, data from the three EHs represent an important validation of the model, since
the Daya Bay data point happens to be in a region where $\alpha$ strongly depends on
$\langle \ethr \cos \theta \rangle$.

\acknowledgments

Daya Bay is supported in part by the Ministry of Science and Technology of China, the U.S. Department of Energy, the Chinese Academy of Sciences, the CAS Center for Excellence in Particle Physics, the National Natural Science Foundation of China, the Guangdong provincial government, the Shenzhen municipal government, the China Guangdong Nuclear Power Group, Key Laboratory of Particle and Radiation Imaging (Tsinghua University), the Ministry of Education, Key Laboratory of Particle Physics and Particle Irradiation (Shandong University), the Ministry of Education, Shanghai Laboratory for Particle Physics and Cosmology, the Research Grants Council of the Hong Kong Special Administrative Region of China, the University Development Fund of The University of Hong Kong, the MOE program for Research of Excellence at National Taiwan University, National Chiao-Tung University, and NSC fund support from Taiwan, the U.S. National Science Foundation, the Alfred~P.~Sloan Foundation, the Ministry of Education, Youth, and Sports of the Czech Republic, the Joint Institute of Nuclear Research in Dubna, Russia, the CNFC-RFBR joint research program, the National Commission of Scientific and Technological Research of Chile, and the Tsinghua University Initiative Scientific Research Program. We acknowledge Yellow River Engineering Consulting Co., Ltd., and China Railway 15th Bureau Group Co., Ltd., for building the underground laboratory. We are grateful for the ongoing cooperation from the China General Nuclear Power Group and China Light and Power Company.

\end{document}